\documentclass[prd,preprint,showpacs]{revtex4}

\usepackage{graphicx}
\usepackage{epstopdf}
\usepackage{latexsym}

\usepackage[center]{subfigure}

\begin{document}

 \newcommand{\bq}{\begin{equation}}
 \newcommand{\eq}{\end{equation}}
 \newcommand{\bqn}{\begin{eqnarray}}
 \newcommand{\eqn}{\end{eqnarray}}
 \newcommand{\nb}{\nonumber}
 \newcommand{\lb}{\label}
\newcommand{\PRL}{Phys. Rev. Lett.}
\newcommand{\PL}{Phys. Lett.}
\newcommand{\PR}{Phys. Rev.}
\newcommand{\CQG}{Class. Quantum Grav.}

\title{Classification of the FRW universe with a  cosmological constant
 and a perfect fluid of the equation of state $p = w\rho$}
\author{Te Ha}
\email{Ha_Ha@baylor.edu}
\author{Yongqing Huang}
\email{Yongqing_Huang@baylor.edu}
\author{Qianyu Ma}
\email{Qianyu_Ma@baylor.edu}
\author{  Kristen D. Pechan}
\email{Kristen_Pechan@baylor.edu}
\author{  Timothy J. Renner}
\email{Timothy_Renner@baylor.edu}
\author{  Zhenbin  Wu}
\email{Zhenbin_Wu@baylor.edu}
\author{  G. A.  Benesh}
\email{Greg_Benesh@baylor.edu}
\author{Anzhong Wang}
\email{anzhong_wang@baylor.edu}


\affiliation{Physics Department, Baylor University, Waco, TX 76798-7316}

\date{\today}

\begin{abstract}

We systematically study the evolution of the Friedmann-Robertson-Walker (FRW) 
universe coupled with a cosmological constant $\Lambda$ and a perfect fluid that 
has the equation of state $p=w\rho$, where $p$ and $\rho$ denote, respectively, 
the pressure and energy density of the fluid, and $w$ is an arbitrary real constant.  
Depending on the specific values of $w,\; \Lambda$, and the curvature $k$ of 
3-dimensional space, we separate all of the solutions into various cases. In each 
case the main properties of the evolution  are given in detail, including the periods 
of deceleration and/or acceleration, and the existence of big bang, big crunch, and 
big rip singularities. In some cases, errors in classification and interpretation appearing 
in standard textbooks have been corrected.  

\end{abstract}


\maketitle

\section{Introduction}
\renewcommand{\theequation}{1.\arabic{equation}} \setcounter{equation}{0}

Recent observations of type Ia supernovas (SN)   reveal the striking discovery that
our universe has lately been in an accelerated expansion phase  \cite{agr98}.
Cross checks from  cosmic microwave background radiation  and large scale
structure all confirm these observations  \cite{Obs}. Such an expansion was predicted neither
by the standard model of particle physics nor by the standard model of cosmology,
and the underlying physics still remains a complete mystery \cite{DEs}.
Since the precise nature and origin of the acceleration have profound
implications, understanding them is one of the major challenges of modern
cosmology. As the Dark Energy Task Force (DETF) stated \cite{DETF}:  ``{\em
Most experts believe that nothing short of a revolution in our understanding
of fundamental physics will be required to achieve a full understanding of
the cosmic acceleration}."

Within the framework of  general relativity (GR),  to account for such an
acceleration requires the introduction of either a tiny positive
cosmological constant  or an exotic component of matter having a very
large negative pressure and interacting weakly with other components of matter.
This invisible component is usually referred to as {\em dark energy}. For a perfect
fluid with the equation of state $w = p/\rho$,  this implies
$w < - 1/3$, where $p$ and $\rho$ denote, respectively, the pressure and energy
density of the fluid. On the other hand, a tiny positive cosmological constant
is entirely consistent with all observations carried out thus far \cite{SCH07}.
However, when the physical origin of the cosmological constant is considered, 
several problems are encountered.  First, the theoretical expectation values exceed observational limits
by $120$ orders of magnitude \cite{wen}.  Second, the corresponding energy density
has only recently become comparable with that of matter--otherwise, galaxies would
not have formed.  Since the energy density of matter is time-dependent, one must 
explain why only {\em now} the two are of the same order.
Third, once the cosmological constant begins to dominate the evolution of the universe,
it dominates forever.  An eternally accelerating universe seems inconsistent
with string/M-Theory because it is endowed with a cosmological event horizon
that prevents the construction of a conventional S-matrix describing particle
interactions \cite{Fish}. Other problems associated with an asymptotic de Sitter universe
in the future were explored in \cite{KS00}.

In view of thse problems, dramatically different models have been proposed,
including quintessence \cite{quit},  DGP branes \cite{DGP00}, and the $f(R)$
models \cite{FR}. For details, see \cite{DEs} and references therein. However,
it is fair to say that so far no convincing model has been proposed.

In this paper, our purpose is two-fold: (i) to provide a complete classification of the FRW models  in terms of classical mechanics; 
and (ii) to correct some errors  appearing
in standard textbooks.  The main idea is simply to cast the dynamical equation in the form of   the conservation of
total energy of a classical particle with mass $m$ moving under the influence of a
potential $V(x)$  \cite{GPS02,szy}, 
\bq 
\lb{1.1} 
\frac{1}{2}m
\dot{x}^{2} + V(x) = E, 
\eq 
where $\dot{x} \equiv dx/dt$ and $E$ is
the total energy of the system. Taking the derivative of the above equation with respect to
$t$, we find that 
\bq 
\lb{1.2} 
m \ddot{x} = - \frac{d V(x)}{dx}. 
\eq
Thus, once the  potential $V(x)$   is known in terms of $x$, it can immediately be determined whether the particle is accelerating or decelerating,
without it being necessary to integrate Eq.(\ref{1.1}) explicitly. In addition, it is simple to determine the range of $x$ allowed by the motion.
Therefore, if the evolution of the universe problem can be expressed in
the above form,  we may use  the  methods of classical mechanics to classify and study all possible solutions.


The rest of the paper is organized as follows: In Sec. II, we
consider the Friedmann equation coupled with a cosmological constant and
a perfect fluid with the equation of state $p = w\rho$ for any given curvature
$k$. After writing it in the form of Eq.(\ref{1.1}), we study  the potential
$V(x)$ case by case, and deduce the main properties of each model of the universe.
In Sec. III, we present our main conclusions.

It should be noted that classification of (non-relativistic)
matter coupled with dark energy was considered recently in \cite{CTS05},
and the corresponding Penrose diagrams were also presented. Similar
considerations were also carried out in a series of papers, and particular
attention was paid to obtaining an effective potential $V(a)$  by fitting observational 
data sets \cite{szy}.  In this paper, we shall not consider the fitting of these models, but simply provide
a complete classification of all such models. Certainly, some of these models
will not be consistent with observations.  

In addition, the method used here is not only applicable to general relativity, but 
also applicable to any theory of gravity, in which the evolution of the universe can be expressed in the form (\ref{1.1}). 
This certainly includes dark energy models where  the equation of state  is a function of the scale factor,
$w = p/\rho = w(a)$. 

In this paper, we use notations and conventions  defined in \cite{d'Inverno}.


\section{Classification of the FRW universe}

\renewcommand{\theequation}{2.\arabic{equation}} \setcounter{equation}{0}

The FRW universe is described by the metric \cite{d'Inverno},
\bq
\lb{2.1}
ds^{2} = dt^{2} - a^{2}(t)\left(\frac{dr^{2}}{1 - kr^{2}}
+ r^{2}\left(d\theta^{2} + \sin^{2}\theta d\phi^{2}\right)\right),
\eq
in the spherically symmetric coordinates $x^{\nu} = \left\{t, r, \theta,
\phi\right\}, \; (\nu = 0, 1, 2, 3)$, where $k$ denotes the curvature of the
three-dimensional space of constant $t$, which can be set to $k = 0, \pm 1$,
without loss of generality. $a(t)$ is the expansion factor of the universe.
It should be noted that Eq.(\ref{2.1}) is invariant
under the translation,
\bq
\lb{2.1a}
t' = t - t_{s},
\eq
where $t_{s}$ is a constant. In the following we shall use this gauge freedom
to fix the origin of the timelike coordinate $t$. The expansion factor $a(t)$
of the universe  is determined through the Einstein field
equations
\bq
\lb{2.2}
R_{\mu\nu} - \frac{1}{2}Rg_{\mu\nu} = \kappa^{2}T_{\mu\nu} + \Lambda g_{\mu\nu},
\eq
where $\kappa^{2} [\equiv 8\pi G/c^{4}]$ is the Einstein coupling constant,
$\Lambda$ denotes the cosmological constant,  and $T_{\mu\nu}$ the energy-momentum
tensor of the matter field(s) filling in the universe. For a perfect fluid,
we have
\bq
\lb{2.3}
T_{\mu\nu} = \left(\rho + p\right) u_{\mu}u_{\nu} - p g_{\mu\nu},
\eq
where $u_{\mu}$ denotes the four-velocity of the fluid and is given by   $u_{\mu} = \delta^{0}_{\mu}$
in the frame of Eq. (\ref{2.1}).
It can be shown \cite{d'Inverno} that the Einstein field equations (\ref{2.2})
for the metric (\ref{2.1}) and energy-momentum tensor (\ref{2.3}), have
only two independent components, which can be cast in the form,
\bqn
\lb{2.4a}
H^{2} &=& \frac{8\pi G}{3} \rho + \frac{1}{3}\Lambda - \frac{k}{a^{2}},\\
\lb{2.4b}
\frac{\ddot{a}}{a} &=& - \frac{4\pi G}{3} \left(\rho + 3p\right)
+ \frac{1}{3}\Lambda,
\eqn
where $H \equiv \dot{a}/a$. Note that in writing the above equation, we have
chosen units such that the speed of light is one. 
On the
other hand, the conservation law of matter fields, $\nabla^{\nu}T_{\mu\nu} = 0$,
yields
\bq
\lb{2.4c}
\dot{\rho} + 3H\left(\rho + p\right) = 0.
\eq
It can be shown that this equation is not independent, and can be obtained
from  Eqs.(\ref{2.4a}) and (\ref{2.4b}). 

Note that we have three unknowns, $a,\; \rho$ and $p$, but only two independent
equations. Thus, to close the system, one more equation is required. Usually
this is given by the equation of state of the matter field. In this paper, we
consider cases in which
\bq
\lb{2.5}
p = w \rho,
\eq
where $w$ is an arbitrary real constant. When $w \ge -1/3$ and $\rho > 0$, the fluid
satisfies all the energy conditions, weak, strong and dominant \cite{HE72}, and usually
is referred to as ``normal" matter. In particular, the fluid with $w = 0$ is a dust, and 
often referred to as matter \cite{SD03}. When $w < -1/3$, the fluid does not satisfy strong
energy condition, and  has a positive contribution to the acceleration of the universe,
as can be seen from Eq. (\ref{2.4b}). This kind of fluid is usually referred to as
dark energy. Since a cosmological constant corresponds to a perfect fluid with
$w = -1$, where the corresponding energy density and pressure are defined by
$\rho_{\Lambda} = - p_{\Lambda} = \Lambda/(8\pi G)$, it is often considered
as a particular dark energy. A fluid with $w < -1$ is called phantom, which does not
satisfy any of the three energy conditions. 

Inserting Eq.(\ref{2.5}) into
Eq.(\ref{2.4c}) and integrating once gives,
\bq
\lb{2.6}
\rho = \rho_{0} \left(\frac{a_{0}}{a}\right)^{3(1+w)},
\eq
where $\rho_{0}$ and $a_{0}$ are integration constants. Since
$\rho_{0}$ represents the energy density when $a = a_{0}$, we
assume that it is strictly positive ($\rho_{0} > 0$).  Without loss of
generality, we can always set $a_{0} = 1$. Then, it can be shown that
the Friedmann equation  (\ref{2.4a}) can be cast in the form of
Eq.(\ref{1.1}) with $m = 1, \; E = 0, \; x(t) = a(t)$, i.e.,
\bq
\lb{2.7}
\frac{1}{2} \dot{a}^{2} + V(a) = 0,
\eq
where
\bq
\lb{2.8}
V(a) = \frac{1}{2}k - \frac{1}{6}\Lambda a^{2}
- \frac{{\cal{C}}}{a^{1+3w}},
\eq
with
\bq
\lb{2.9}
{\cal{C}} \equiv \frac{4\pi G\rho_{0}}{3} > 0.
\eq
It should be noted that the equivalency of the Friedmann equation  (\ref{2.7}) 
with the conservation law of mechanical energy Eq.(\ref{1.1}) is rather a formality.
In particular, energy   in GR is not well-defined, while a force is no longer a 
vector. Therefore, although we can still use these terms to describe the motion of the 
universe in the following discussions, one should   always  keep this  in mind. 
 
When $w = 0$ the problem reduces to the one treated in \cite{d'Inverno}.
To study the problem further, we consider separately the cases $k = 0, \pm 1$.

\subsection{$k = 0$}

When $k = 0$, Eq.(\ref{2.8}) reduces to
\bq
\lb{2.8a}
V(a) =  - \frac{1}{6}\Lambda a^{2}
- \frac{{\cal{C}}}{a^{1+3w}}.
\eq
It is convenient to consider the cases    $\Lambda > 0, \; \Lambda = 0$, and
$ \Lambda < 0$ separately.

\subsubsection{$\Lambda > 0$}

When $\Lambda > 0$, Eq.(\ref{2.8a}) can be written as
\bq
\lb{2.8ab}
V(a) =  - \frac{1}{6}\Lambda a^{2}\left( 1+ \frac{\tilde{{\cal{C}}}}{a^{3(1+w)}}\right),
\eq
where $\tilde{{\cal{C}}} \equiv 6{\cal{C}}/\Lambda > 0$. Thus, in this case $V(a)$ is always
non-positive. Depending on the
value of $w$, the evolution of the universe can differ significantly. Thus, we further
distinguish the following sub-cases:
\bqn
\lb{2.10}
& & (1) \; w > - \frac{1}{3},\;\;\; (2) \; w = - \frac{1}{3},\;\;\;
(3) \; - 1 < w < - \frac{1}{3},\nb\\
& & (4) \; w  = -1 ,\;\;\; (5) \;  w < - 1.
\eqn

\begin{figure}
\includegraphics[width=\columnwidth]{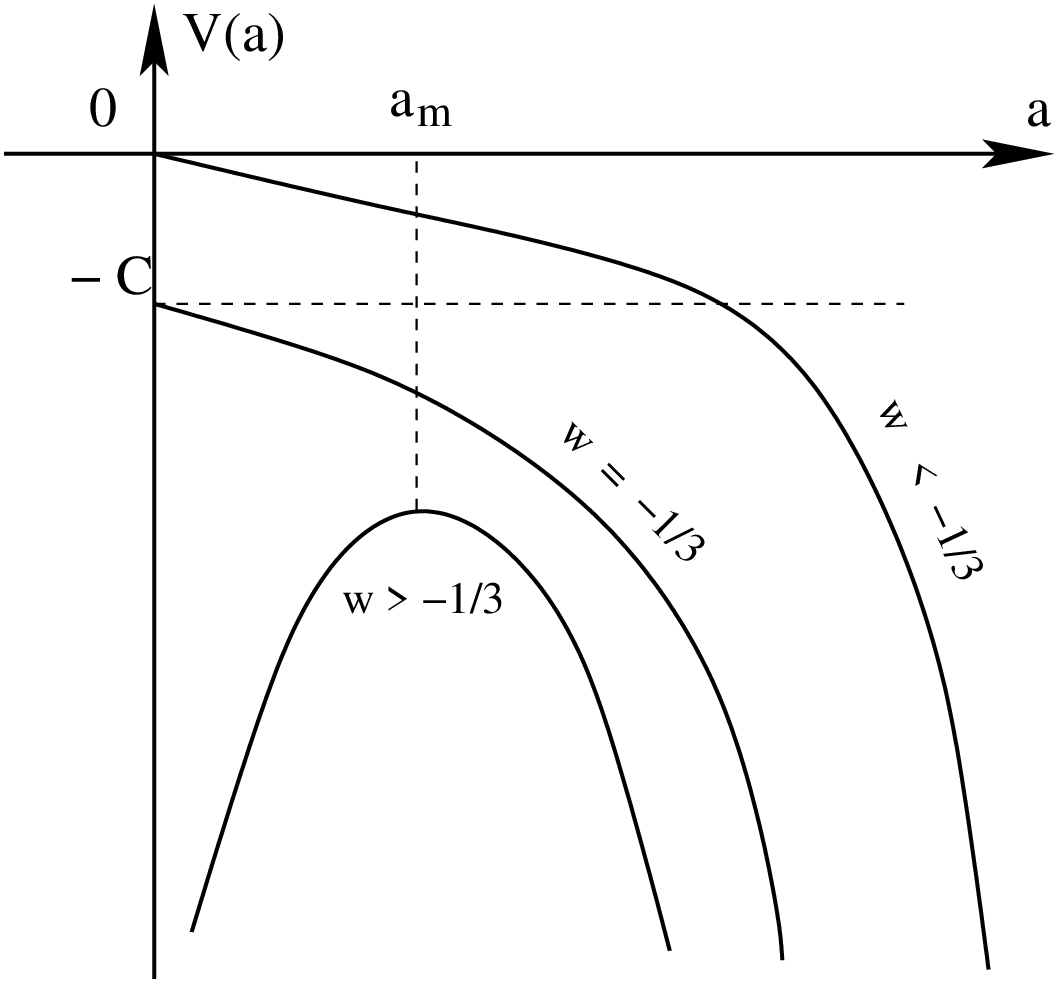}
\caption{The potential given by Eq.(\ref{2.8ab})   for $k = 0$ and $\Lambda > 0$.}
\label{fig1}
\end{figure}

{\bf Case A.1.1) $w > - \frac{1}{3}$}: Then, we find that  $V(a) \rightarrow - \infty$ for both $a = 0$
and $a \rightarrow \infty$. It also has a maximum at $a = a_{m} \equiv \left(3\left(1+3w\right)
{\cal{C}}/\Lambda\right)^{1/(3(1+w))}$, for which
\bq
\lb{2.11}
\ddot{a}  = \cases{ < 0, & $ a < a_{m}$, \cr
 = 0, & $ a = a_{m}$, \cr
  > 0, & $ a > a_{m}$.}
\eq
Fig. \ref{fig1} schematically shows the potential, represented by the bottom line. Therefore, in
this case, the evolution of the universe is dominated by matter in early
times and $a(t) \propto t^{2/[3(1+w)]}$, for which $\ddot{a} < 0$. As
the universe expands to $a = a_{m}$, it reaches the turning point, 
after which the expansion begins accelerating, i.e., $\ddot{a} > 0$ for  $ a > a_{m}$. The universe is asymptotically
de Sitter, $a(t) \propto e^{\sqrt{\Lambda/3} t}$. 

{\bf Case A.1.2) $w = - \frac{1}{3}$}: In this case, we find
\bqn
\lb{2.12}
V(a)  &=&  - \frac{1}{6}\Lambda a^{2} - {\cal{C}} \nb\\
&=& \cases{ - {\cal{C}}, & $ a =  0$, \cr
 -\infty,   & $ a \rightarrow \infty$, \cr}\nb\\
\ddot{a} &=& - \frac{dV(a)}{da} = \frac{1}{3}\Lambda a \ge 0.
\eqn
The middle line of Fig. \ref{fig1} represents the potential for this case. The universe is always
accelerating after the initial moment when  $a = 0$. On the other hand, from Eq.(\ref{2.6})
we find that $\rho \propto a^{-2} \rightarrow \infty$ as $a \rightarrow 0$;  that is,
a big bang singularity still occurs at $a = 0$.

{\bf Case A.1.3) $-1 < w < - \frac{1}{3}$}:
In this case, the potential is represented by the upper line in Fig. \ref{fig1}, and
\bq
\lb{2.13}
\ddot{a} = - \frac{dV(a)}{da} = \frac{1}{3}\Lambda a
    + \left(3|w| - 1\right){\cal{C}} a^{3|w| - 2} \ge 0,
\eq
where the equality holds only at $a = 0$ for $w \not= -2/3$. Thus, in this case the universe
is always accelerating. Note that in the present case $a = 0$ still represents a big bang
singularity, as can be seen from Eq.(\ref{2.6}). It is also interesting to note that at
$a = 0$, we have $\dot{a} = \ddot{a} = 0$, except for  $w = -2/3$. Thus, when $w \not= -2/3$
the point $a = 0$ is a stationary point. However, it is not stable, and any perturbation
will cause the universe to expand. When  $w = -2/3$, we have $\dot{a}(a=0) = 0$ and $\ddot{a}(a=0)
 = \left(3|w| - 1\right){\cal{C}} > 0$. Therefore, $a = 0$ is not stationary in the latter case,
 and the positive force causes the universe to expand automatically.

{\bf Case A.1.4) $ w = -1$}: In this case, matter acts as a vacuum energy,
and the potential is given by
\bq
\lb{2.14}
V(a) =  - \frac{1}{6} \Lambda_{eff} a^{2},
\eq
where $\Lambda_{eff} \equiv \Lambda + 6{\cal{C}}$. Therefore,  the universe is de
Sitter, and
\bq
\lb{2.14b}
a(t) = e^{\sqrt{\Lambda_{eff}/3}\; (t - t_{0})}.
\eq
Recall that the de
Sitter space is free of any kind spacetime singularities at $a = 0$, as well as
at $a = \infty$ \cite{d'Inverno}.

{\bf Case A.1.5) $ w < -1$}: In this case, the behavior of the
potential $V(a)$ and $a(t)$ are similar to the case $ - 1 < w <
-1/3$, except  that now $\rho \propto a^{3(|w| - 1)}$ is
not singular at $a = 0$. However, there is a singularity at $a = \infty$, which is
usually referred to as a big rip singularity. At $a = 0$ we have $\dot{a} =
\ddot{a} = 0$; thus, this point  also represents an
unstable stationary point.

Finally, we note that  for all cases with $k = 0$, the corresponding Friedmann
equation can be integrated explicitly, and the corresponding solutions are given by,
\bq
\lb{2.15}
a(t) =\left\{\left(\frac{6{\cal{C}}}{\Lambda}\right)^{1/2} \sinh\left[(1+w)
\sqrt{\frac{3\Lambda}{4}}\left(t - t_{s}\right)\right]\right\}^{\frac{2}{3(1+w)}},
\eq
for $w \not= -1$,   where $t_{s}$ is given by
\bq
\lb{2.16}
t_{s} = t_{0} - \frac{1}{1+w}\sqrt{\frac{4}{3\Lambda}}\sinh^{-1}
\left(\sqrt{\frac{\Lambda}{6{\cal{C}}}}\right),
\eq
so that $a(t = t_{0}) = 1$. For $w > -1$, without loss of generality, we  
use the gauge freedom of Eq.(\ref{2.1a}) to set $t_{s} = 0$, so that the big bang
singularity occurs at $t = 0$. This will be the case for the rest of this paper.
 
When $w = -1$, the solution is de Sitter, given by
Eq.(\ref{2.14b}), which is free of all spacetime singularities. The
solution is valid for all $t$, that is, $t \in \left(-\infty, \infty\right)$. For the
cases in which $w < -1$, there is no translation,
so that the big rip singularity occurs exactly at $t = t_{s}$.
Note that when $w < -1$, solutions of Eq.(\ref{2.15}) are valid
only when   $t \in ( - \infty, t_{s})$, for which
\bqn
\lb{2.17}
a(t) &=& \left\{\sqrt{\frac{6{\cal{C}}}{\Lambda}} \sinh\left[\sqrt{\frac{3\Lambda}{4}}
\left(|w| -1\right)
\left(t_{s} - t\right)\right]\right\}^{- \frac{2}{3(|w| -1)}}\nb\\
&=& \cases{\infty, & $t = t_{s}$,\cr
  1, & $t = t_{0}$,\cr
  0, & $t = - \infty$,\cr} \; (w < -1).
\eqn
Since $\rho \propto a^{3(|w| -1)} $ spacetime is indeed
not singular at $a(t= -\infty) =0$, but is singular at $a(t = t_{s}) = \infty$.

\begin{figure}
\includegraphics[width=\columnwidth]{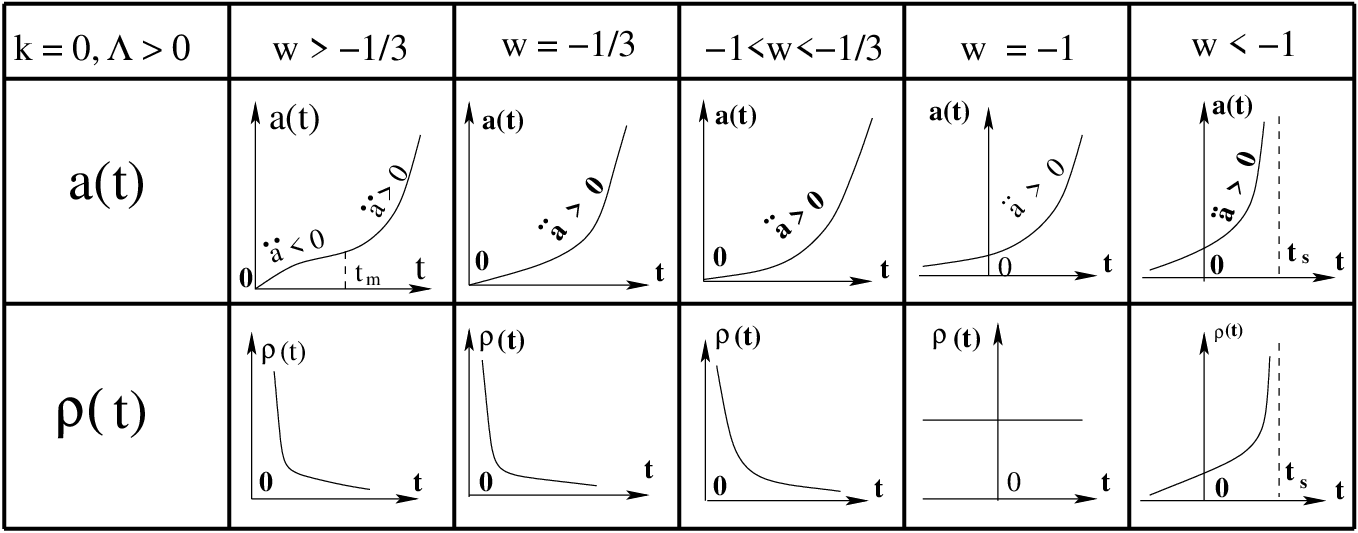}
\caption{The expansion factor $a(t)$, the acceleration
$\ddot{a}(t)$, and the energy density $\rho(t)$ for  $k = 0$ and $
\Lambda > 0$. A big bang singularity occurs at $t = 0$
for $w > -1$. It is de Sitter for $w = -1$, which is free of any
kind of spacetime singularities. When $w < -1$,  a big rip
singularity occurs at $t = t_{s}$, in which  
$a\left(t_{s}\right) = \infty = \rho\left(t_{s}\right)$. }
\label{figA}
\end{figure}

In Fig. \ref{figA} we summarize the main properties of the solutions for $k = 0$ and
$\Lambda > 0$ for different values of $w$.

\subsubsection{$\Lambda = 0$}

\begin{figure}
\includegraphics[width=\columnwidth]{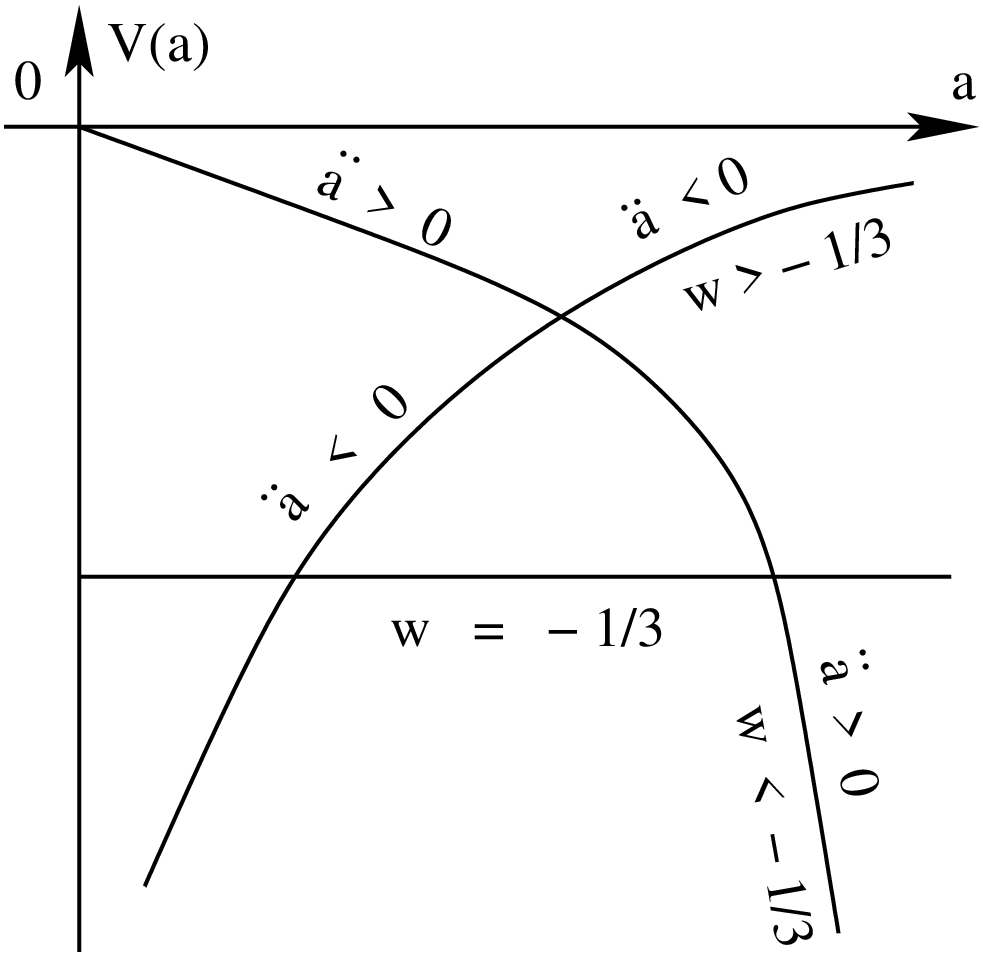}
\caption{The potential given by Eq.(\ref{2.8ab})   for $k = 0$ and $\Lambda = 0$.}
\label{fig2}
\end{figure}

In this case, we have
\bqn
\lb{2.18}
V(a) &=& - \frac{{\cal{C}}}{a^{1+3w}} \le 0,\nb\\
\ddot{a} &=& - (3w + 1) \frac{{\cal{C}}}{a^{2+3w}}\nb\\
&=& \cases{ < 0, & $w> -1/3$,\cr
 = 0, & $w= -1/3$,\cr
  > 0, & $w< -1/3$.\cr}
\eqn
Fig. \ref{fig2} shows the potential $V(a)$. As in the previous case, we may integrate the Friedmann equation (\ref{2.4a}) to obtain explicit solutions
for $a(t)$ and $\rho(t)$,
\bqn
\lb{2.19}
a(t) &=& \cases{\left[3(1+w)\sqrt{\frac{{\cal{C}}}{2}}
        \left(t - t_{s}\right)\right]^{\frac{2}{3(1+w)}}, & $w \not= -1$,\cr
    e^{\sqrt{2{\cal{C}}}(t - t_{0})}, & $w = -1$,\cr}\nb\\
\rho(t) &=& \cases{\frac{\tilde{\rho}_{0}}{(t - t_{s})^{2}}, & $w \not= -1$,\cr
\rho_{0}, & $w = -1$,\cr},
\eqn
where $\tilde{\rho}_{0} \equiv 2\rho_{0}/[9(1+w)^{2}{\cal{C}}]$, and
\bq
\lb{2.20}
t_{s} = t_{0} - \frac{1}{1+w}\sqrt{\frac{2}{9{\cal{C}}}}.
\eq

When $w < -1$, Eq.(\ref{2.19}) shows that $a(t)$ remains real and positive when $t \in (-\infty, t_{s})$. As $t \rightarrow - \infty$, both $a(t)$
and $\rho(t)$ vanish; however, when  $t \rightarrow t_{s}$ each becomes unbounded--that is, a big rip singularity is developed there.

\begin{figure}
\includegraphics[width=\columnwidth]{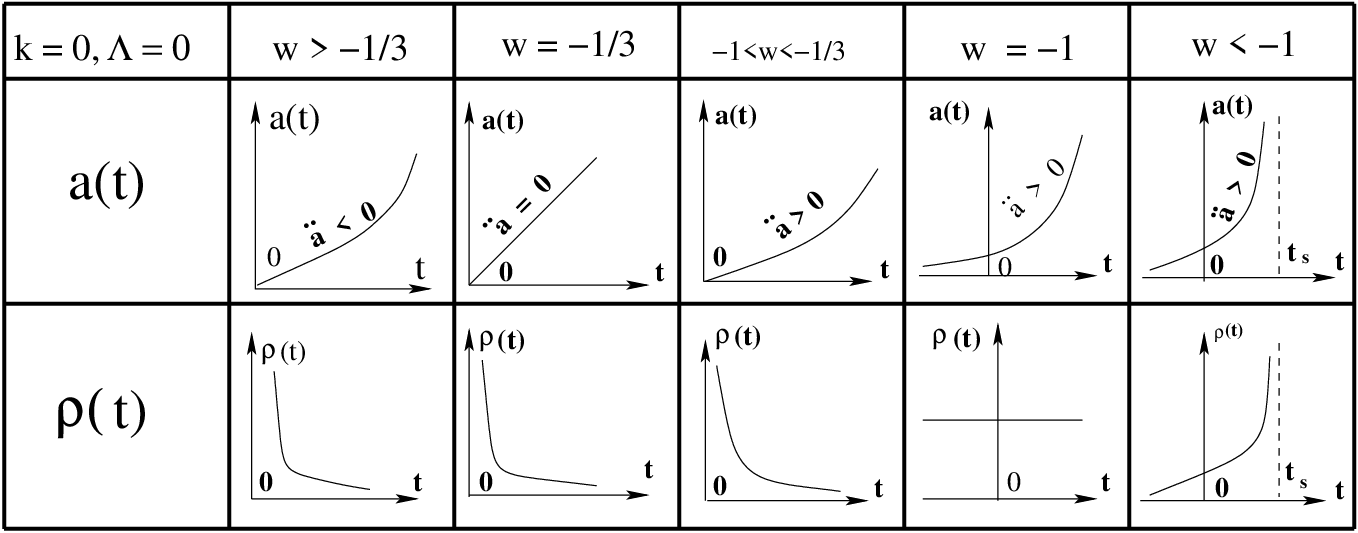}
\caption{The expansion factor $a(t)$, the acceleration
$\ddot{a}(t)$, and the energy density $\rho(t)$ for  $k = 0$ and $
\Lambda = 0$. There is a big bang singularity at $a = 0$ for all 
cases with $w > -1$. The spacetime is de Sitter for $w = -1$. When
$w < -1$,  a big rip singularity is developed at $t = t_{s}$, at
which we have $a\left(t_{s}\right)  = \rho\left(t_{s}\right) =
\infty$. } \label{figB}
\end{figure}

In Fig. \ref{figB} we summarize the main properties of the $k = 0$ and
$\Lambda = 0$ solutions for different values of $w$.

\subsubsection{$\Lambda < 0$}

\begin{figure}
\includegraphics[width=\columnwidth]{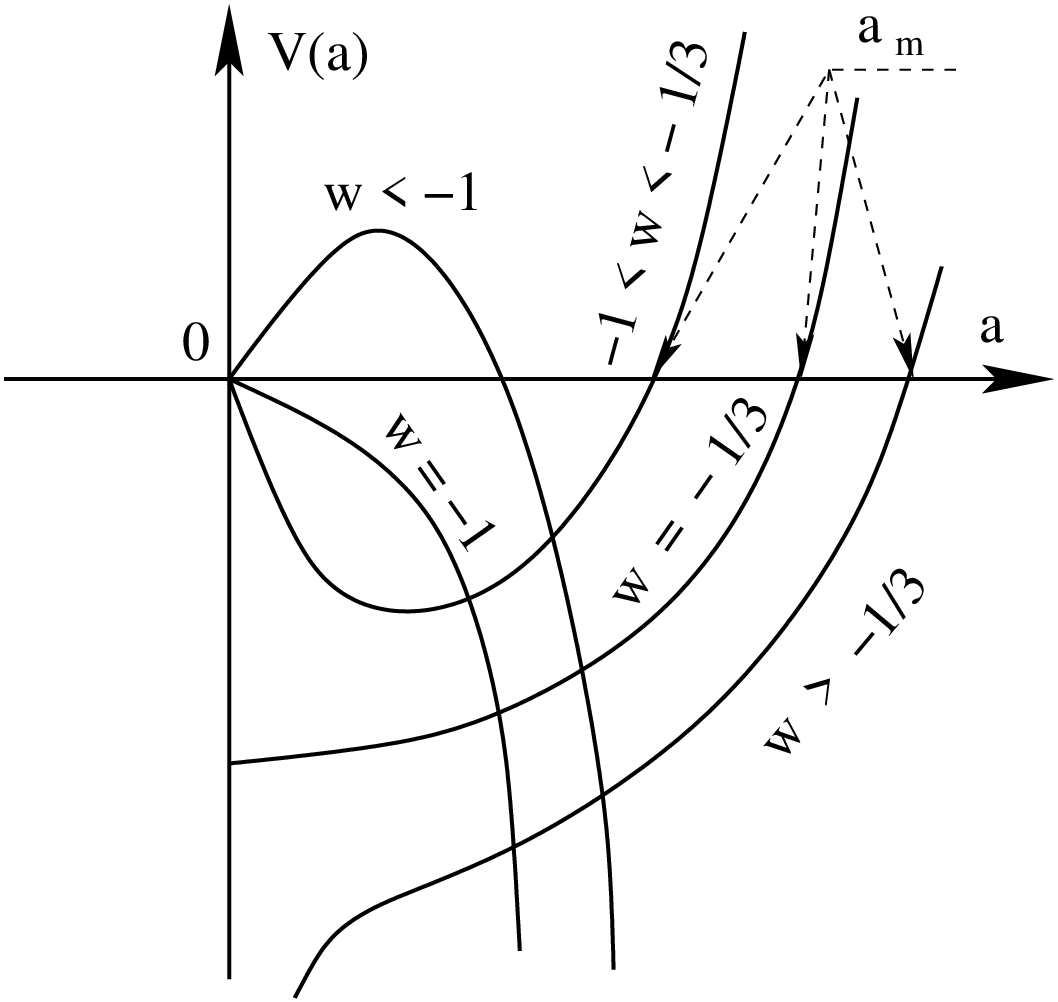}
\caption{The potential given by Eq.(\ref{2.21})   for $k = 0$ and $\Lambda < 0$.}
\label{fig3}
\end{figure}

In this case, we have
\bq
\lb{2.21}
V(a) =   \frac{1}{6}\left|\Lambda\right| a^{2}
- \frac{{\cal{C}}}{a^{1+3w}},
\eq
which is illustrated in Fig. \ref{fig3} for various values of $w$. It can be
shown that the corresponding expansion factor and energy density are given,
respectively, by
\bqn
\lb{2.22a}
a(t) &=& \left\{\sqrt{\frac{6{\cal{C}}}{|\Lambda|}}
\sin\left[\sqrt{\frac{3|\Lambda|}{4}}(1+w)\left(t-t_{s}\right)\right]\right\}
  ^{\frac{2}{3(1+w)}}, \nb\\
\rho(t) &=& \frac{\tilde{\rho}_{0}}
{\sin^{2}\left(\sqrt{\frac{3|\Lambda|}{4}}(1+w)\left(t-t_{s}\right)\right)},
\eqn
for $w \not= -1$, and
\bqn
\lb{2.22b}
a(t) &=& e^{\sqrt{\Lambda_{eff}/3}\left(t - t_{0}\right)},\nb\\
\rho(t) &=& \rho_{0},
\eqn
for $w = -1$, where $\Lambda_{eff} = |\Lambda| - 6{\cal{C}} > 0$.
Also,
\bqn
\lb{2.22c}
t_{s} &\equiv& t_{0} - \frac{\sqrt{4/(3|\Lambda|)}}{1+w}\sin^{-1}
\left(\sqrt{\frac{|\Lambda|}{6{\cal{C}}}}\right),\nb\\
\tilde{\rho}_{0} &\equiv& \frac{|\Lambda|\rho_{0}}{6{\cal{C}}}.
\eqn

As in the previous cases, we use gauge freedom (\ref{2.1a}) to set $t_{s} = 0$ for $w > -1$, while keeping $t_{s}$ unchanged for $w \le -1$.

\begin{figure}
\includegraphics[width=\columnwidth]{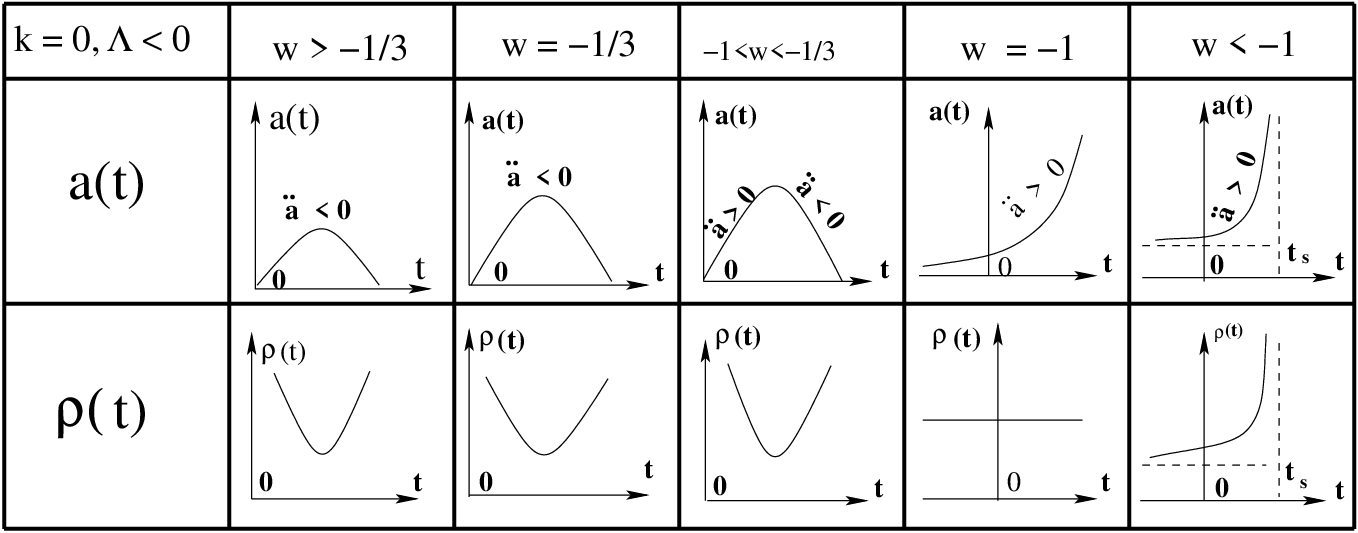}
\caption{The expansion factor $a(t)$, the acceleration $\ddot{a}(t)$, and the energy
density $\rho(t)$ for  $k = 0$ and $ \Lambda < 0$. The spacetime  is singular
at $a = 0$ for all cases with $w > -1$, (a big bang singularity). It is de Sitter
for $w = -1$. When $w < -1$,  a big rip singularity is developed at $t = t_{s}$,
at which we have $a\left(t_{s}\right)  = \rho\left(t_{s}\right) = \infty$. }
\label{figC}
\end{figure}

{\bf Case A.3.1) $w > - \frac{1}{3}$}: In this case, as shown in Fig. \ref{fig3}, $V\left(a> a_{m}\right) > 0$; so, the motion for $a> a_{m}$ is forbidden. If the
universe starts expanding from the big bang when $a = 0$, after expanding to its
maximal radius $a_{m}$ it will start collapsing until a big crunch singularity
is developed at $t = 2t_{m}$. This is illustrated in Figs. \ref{fig3} and \ref{figC}, where
$a_{m} = a\left(t_{m}\right)$ is given by
\bq
\lb{2.23}
a_{m} = \left(\frac{6{\cal{C}}}{|\Lambda|}\right)^{\frac{1}{3(1+w)}}.
\eq
During the process,  the universe is always decelerating,
\bq
\lb{2.24}
\ddot{a} = - \frac{dV(a)}{da} < 0,
\eq
as can be seen from Fig. \ref{fig3}.

{\bf Case A.3.2) $w = - \frac{1}{3}$}: In this case, we find
\bqn
\lb{2.25}
V(a)  &=&   \frac{1}{6}\left|\Lambda\right| a^{2} - {\cal{C}}
= \cases{ - {\cal{C}}, & $ a =  0$, \cr
 \infty,   & $ a \rightarrow \infty$, \cr}\nb\\
\ddot{a} &=& - \frac{dV(a)}{da} = - \frac{1}{3}|\Lambda| a \le 0.
\eqn Therefore, as in the last case, the universe  expands from
the big bang singularity at $a = 0$ until its maximal radius
$a_{m}$, given by Eq.(\ref{2.23}) with $w = -1/3$ is reached. It then starts
to collapse until a big crunch singularity is formed at $t =
2t_{m}$.

{\bf Case A.3.3) $-1 < w < - \frac{1}{3}$}: In this case, from Fig. \ref{fig3}
we can see that  $V(a = 0) = 0 = V\left(a_{m}\right)$, and the motion is also
restricted to $a \le a_{m}$. However, there is a fundamental difference between this
case and the previous two cases: the potential $V(a)$ has a minimum at $a = a_{min}$,
at which $dV\left(a_{min}\right)/da = 0$. The universe accelerates initially. However, 
after expanding to $a = a_{min}$, it starts to decelerate
until $a = a_{m}$, at which point the expansion velocity becomes zero. Afterward
it will begin collapsing, until a big crunch singularity is developed at $a = 0$,
as shown in Fig. \ref{figC}.

{\bf Case A.3.4) $ w = -1$}: In this case, we have
\bq
\lb{2.26}
V(a) =  - \frac{1}{6} \left|\Lambda\right| a^{2}\left(\frac{6{\cal{C}}}
{\left|\Lambda\right|} - 1\right).
\eq
Therefore, there is a solution only when $\Lambda_{eff} > 0$, for which
the universe is de Sitter, and
\bq
\lb{2.27}
a(t) = e^{\sqrt{\Lambda_{eff}/3}\; (t - t_{0})},
\eq
where $\Lambda_{eff} \equiv 6{\cal{C}} - \left|\Lambda\right|$.

{\bf Case A.3.5) $ w < -1$}: In this case,  there is a minimum $a_{min}$ for
which    $V\left(a< a_{min}\right) \ge 0$. Therefore, in contrast to the
previous case, the motion of the universe is restricted to $a \ge a_{min}$.
As shown in Fig. \ref{figC}, the universe starts to expand from $a = a_{min}$.  Within a finite time $a\left(t_{s}\right)
= \infty$, and a big rip singularity is formed.

\subsection{$k = 1$}

In this case, the potential given by Eq.(\ref{2.8}) can be written
as
\bq
\lb{2.28}
V(a) = \frac{1}{2} - \frac{1}{6}\Lambda a^{2}\left[1
 + \left(\frac{\tilde{\cal{C}}}{a}\right)^{3(1+w)}\right],
\eq
where $\tilde{\cal{C}} \equiv \left(6{\cal{C}}/\Lambda\right)^{\frac{1}{3(1+w)}}$.
As in the   $k = 0$ case, we  distinguish the three cases,
$\Lambda > 0, \; \Lambda = 0$ and $\Lambda < 0$.

\subsubsection{$\Lambda > 0$}

When $\Lambda > 0$,   it is convenient to further divide into
the five sub-cases listed in Eq.(\ref{2.10}).

{\bf Case B.1.1) $\; w  > - \frac{1}{3}$}: In this case, it can be shown
that for any given $w$ and $\rho_{0}$ there always exist a critical
value $\Lambda_{c}$ and radius $a_{m}$ satisfying the conditions,
\bq
\lb{2.28a}
V\left(a_{m}, w, \rho_{0}, \Lambda_{c}\right) = 0,\;\;\;
V'\left(a_{m}, w, \rho_{0}, \Lambda_{c}\right) = 0,
\eq
where a prime denotes ordinary differentiation with respect to $a$. 
The solutions for these conditions are
\bqn
\lb{2.28b}
a_{m} &=& \left[3\left(1+w\right){\cal{C}}\right]^{\frac{1}{1+3w}},\nb\\
\Lambda_{c} &=& \left(\frac{1+3w}{1+w}\right)
\left[3\left(1+w\right){\cal{C}}\right]^{-\frac{2}{ 1+3w}}.
\eqn
As  shown below, the solutions with $\Lambda > \Lambda_{c}$
have quite different properties from the ones with $\Lambda < \Lambda_{c}$.
Therefore, we further distinguish the three cases, $\Lambda > \Lambda_{c}, \; \Lambda = \Lambda_{c}$ and
$\Lambda < \Lambda_{c}$.

\begin{figure}
\includegraphics[width=\columnwidth]{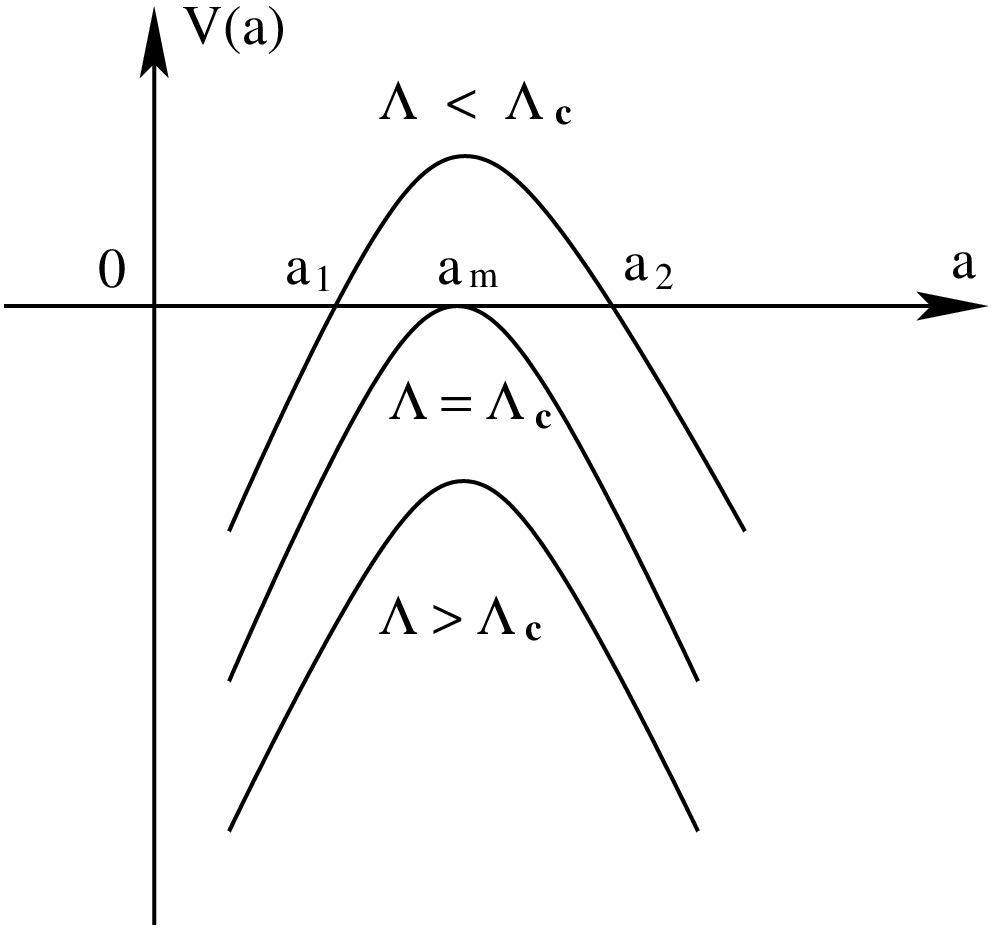}
\caption{The potential given by Eq.(\ref{2.28})
for $k = 1, \; w > - \frac{1}{3}$ and $ \Lambda > 0$,
where $\Lambda_{c} = \Lambda_{c}\left(w, \rho_{0}\right)$.}
\label{fig4}
\end{figure}

\begin{figure}
\includegraphics[width=\columnwidth]{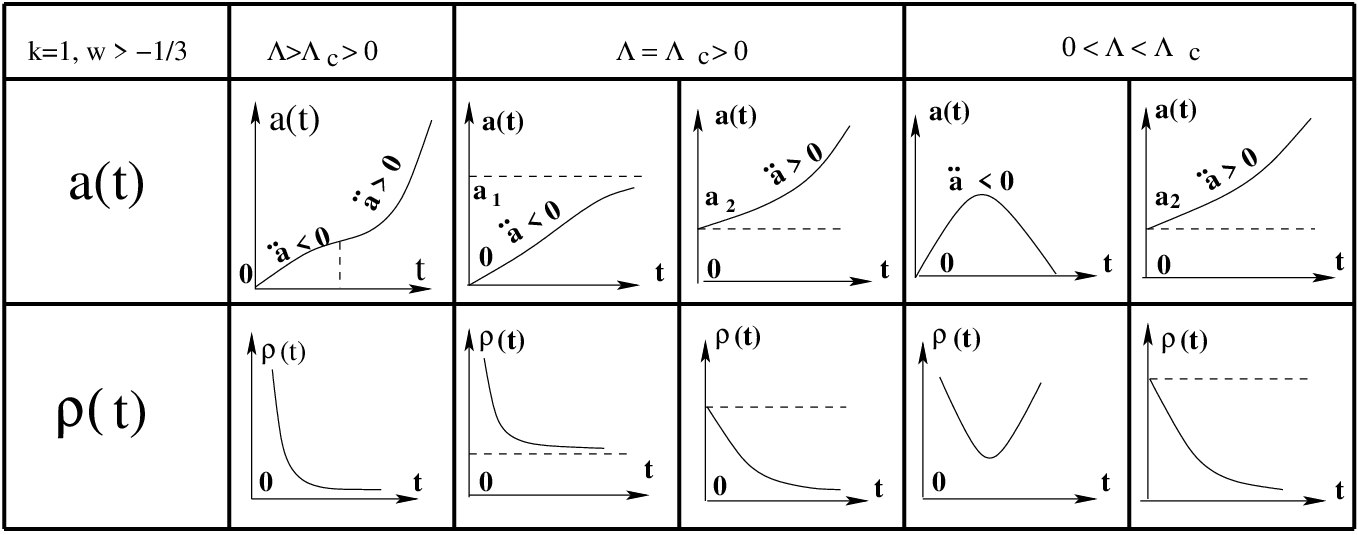}
\caption{The expansion factor $a(t)$, the acceleration $\ddot{a}(t)$, and the energy
density $\rho(t)$ for  $k = 1, \; w > - \frac{1}{3}$ and $ \Lambda > 0$, where
$\Lambda_{c}$ is given by Eq.(\ref{2.28b}).   A big bang singularity
occurs at $a= 0$ in all cases with $\Lambda \ge \Lambda_{c}$. In the first
sub-case of $0 < \Lambda < \Lambda_{c}$, both big bang and big crunch singularities
occur, while in the second sub-case,  spacetime is free of any kind of spacetime
singularities. }
\label{figD}
\end{figure}

{\bf Case B.1.1.1) $\; w  > - \frac{1}{3}, \; \Lambda > \Lambda_{c}$}:
In this case,  the potential
$V(a)$ is always negative for any given $a$, as shown in Fig. \ref{fig4}. Therefore,
the corresponding solutions have no turning points. If the universe initially
starts to expand from a big bang singularity at $a = 0$, it will expand forever,
as shown by  Fig. \ref{figD}. However, the potential has a maximum at $a = a_{m}$,
for which we have
\bq
\lb{2.29}
\ddot{a} = \cases{ < 0, & $ a < a_{m}$,\cr
 = 0, & $ a = a_{m}$,\cr
  > 0, & $ a > a_{m}$,\cr}
\eq
that is, the universe is initially decelerating. Once it expands to $a = a_{m}$,
the expansion begins to accelerate.

{\bf Case B.1.1.2) $\; w  > - \frac{1}{3}, \; \Lambda =
\Lambda_{c} $}: In this case,  there exists a static point $a_{m}$,
at which we have $V\left(a_{m}\right) = V'\left(a_{m}\right) = 0$,
as seen in Fig. \ref{fig4}. Therefore, if the universe
starts to expand from the big bang at $a = 0$, it will expand until
$a = a_{m}$. The universe is decelerating during this period. At the point $a = a_{m}$, 
the universe becomes static, since $\dot{a} = 0 = \ddot{a}$.   However,
this point is not stable, and with small perturbations, the universe will
either collapse until a singularity is developed at $a  = 0$ or
expand forever with $\ddot{a} > 0$. If the universe initially has an acceleration $a
= a_{i} > a_{m}$, then as shown in Fig. \ref{fig4}, it will
expand forever. Since $V'(a)$ is always negative, the universe
in this region always accelerates.

{\bf Case B.1.1.3) $\; w  > - \frac{1}{3}, \; 0 < \Lambda < \Lambda_{c}$}:
In this case, $V(a) = 0$ has
two real and positive roots, $a_{1}$ and $a_{2}$, as shown in Fig. \ref{fig4}.
Without loss of generality, we assume that $a_{2} > a_{1}$. Since $V(a) > 0$
for $a \in \left(a_{1}, a_{2}\right)$, the motion is forbidden in this region.
As in the previous case, depending on the initial conditions, the universe can
evolve very differently. In particular, if it starts to expand from the
big bang singularity at $a = 0$, it will expand until $a = a_{1}$, 
where $\dot{a} = 0$ and $\ddot{a} < 0$. Since $\ddot{a} < 0$ at this point,  the
universe will start collapsing, until a big crunch singularity is
developed at $a = 0$. If the universe starts to expand at $a_{i} \ge a_{2}$, it
will expand forever. In the latter case, the universe is always accelerating, as
can be seen in Fig. \ref{fig4}.

\begin{figure}
\includegraphics[width=\columnwidth]{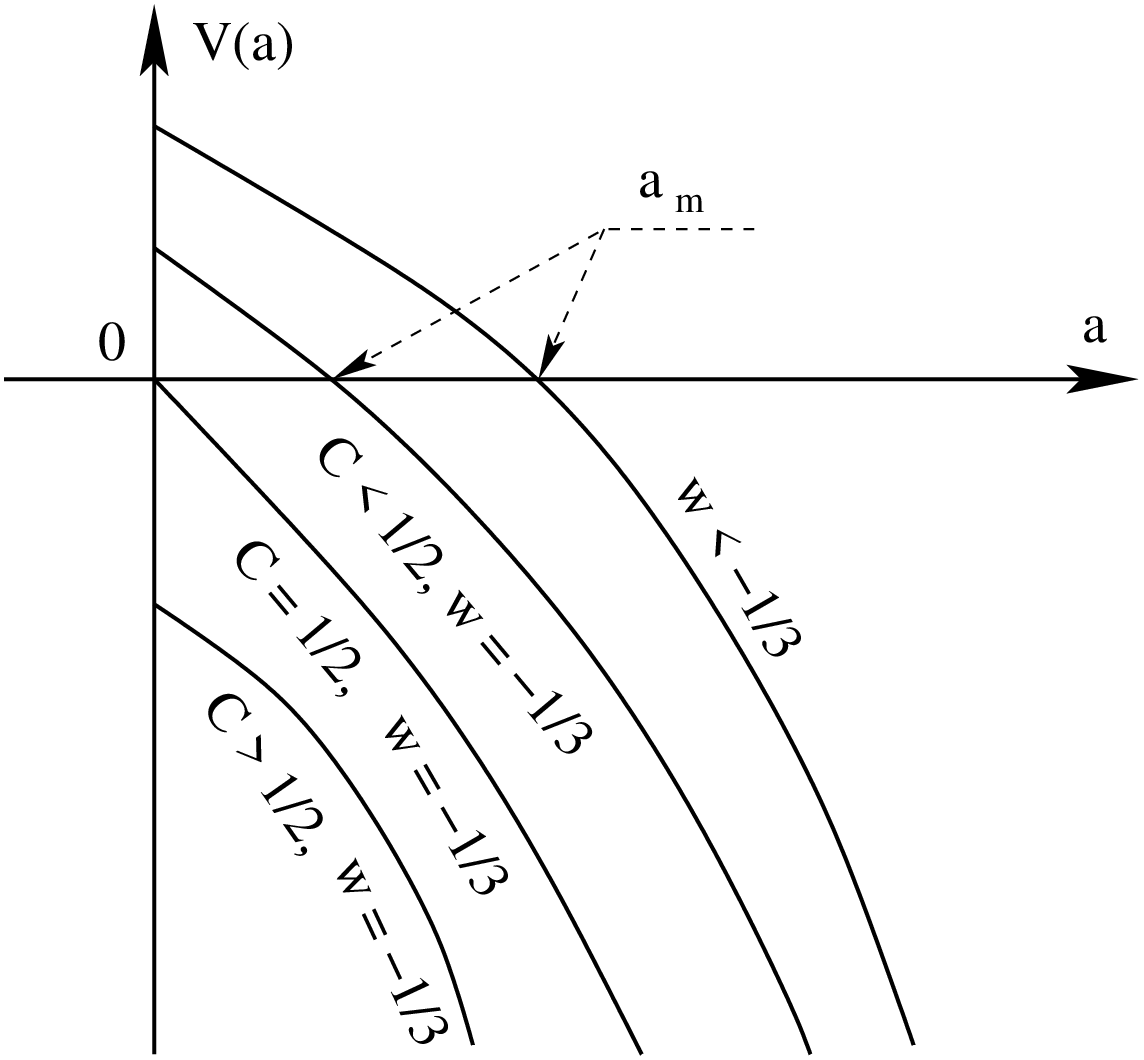}
\caption{The potential given by Eq.(\ref{2.30})  for
$k = 1, \; w \le - \frac{1}{3}$ and $ \Lambda > 0$.}
\label{fig5}
\end{figure}

\begin{figure}
\includegraphics[width=\columnwidth]{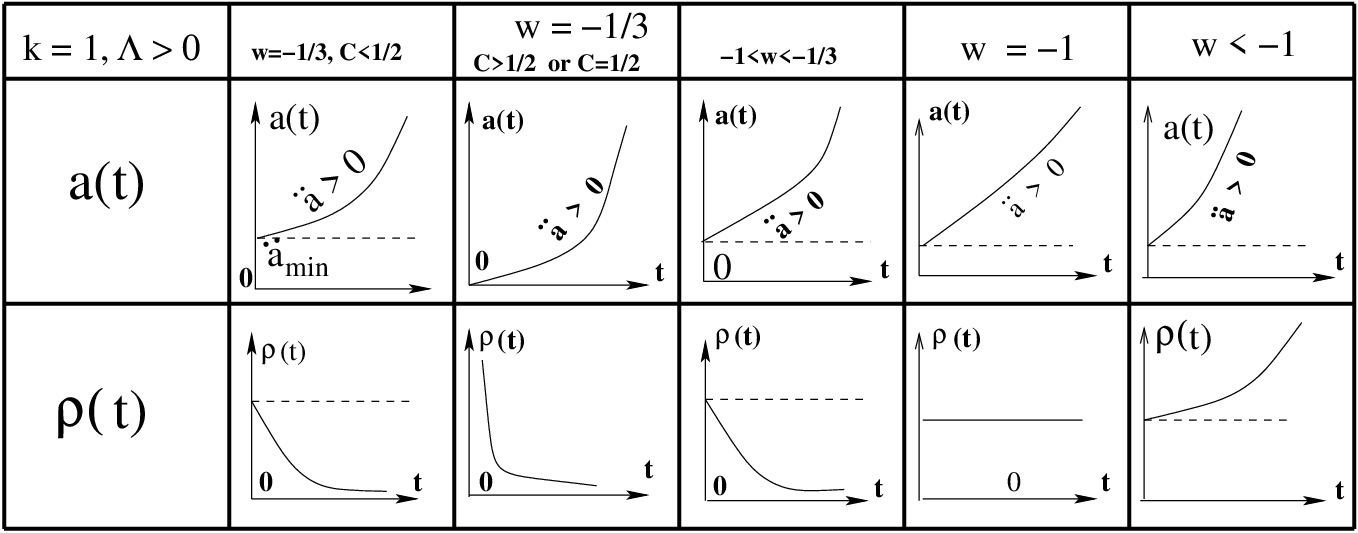}
\caption{The expansion factor $a(t)$, the acceleration $\ddot{a}(t)$, and the energy
density $\rho(t)$ for  $k = 1, \; w \le - \frac{1}{3}$ and $ \Lambda > 0$.
A big bang   singularity occurs   only in the case   $w = -1/3$ and ${\cal{C}} \ge 1/2$.
In the case $w < -1$, a big rip  singularity occurs at $a= \infty$. }
\label{figE}
\end{figure}

{\bf Case B.1.2) $\; w  = - \frac{1}{3}$}: In this case, we have
\bq
\lb{2.30}
V(a) = \frac{1}{2}\left(1 - {2\cal{C}}\right) -  \frac{1}{6}\Lambda a^{2}.
\eq
Thus, the motion of the
universe will differ depending on the value of ${\cal{C}}\left(\rho_{0}\right)$. In particular, when ${\cal{C}}\left(\rho_{0}\right) < 1/2$,
there exists a minimal $a_{min}$, for which $V\left(a < a_{min}\right) > 0$; that
is, the motion in the region $0 < a < a_{min}$ is forbidden, as shown in Fig.
\ref{fig5}. When ${\cal{C}}\left(\rho_{0}\right) \ge 1/2$, the universe starts to
expand from the big bang singularity at $a = 0$. In all cases we have $V'(a) < 0$,
so that the universe is always accelerating [cf. Fig. \ref{figE}].

{\bf Case B.1.3) $\; - 1 < w  < - \frac{1}{3}$}: In this case, we find that
$V'(a)$ is strictly negative for any $a \ge 0$ with $V(0) = 1/2$. Therefore, similar
to the case $w = -1/3$ and ${\cal{C}} < 1/2$, there exists a minimal $a_{min}$, for
which $V\left(a < a_{min}\right) > 0$, and the motion in the region $0 < a < a_{min}$
is forbidden, as shown in Fig. \ref{fig5}. Thus, in the present case,  the universe
starts to expand from a radius $a_{i} \ge a_{min}$ until $a = \infty$ without a turning point.
Since $V'(a) < 0$ for any $a \ge a_{min}$,  the universe is always
accelerating. No singularities develop: not
big bang, big crunch,  nor big rip, as shown in Fig. \ref{figE}.

{\bf Case B.1.4) $\;  w  = -1$}: In this case, the potential is simply parabolic,
\bq
\lb{2.31}
V(a) = \frac{1}{2}  -  \frac{1}{6}\left(\Lambda + 6{\cal{C}}\right) a^{2},
\eq
schematically shown by the top curve in Fig. \ref{fig5}. As a result, the motion is
similar to the last case, except that now $\rho = \rho_{0}$.

{\bf Case B.1.5) $\;   w  < - 1$}: In this case, we also have $V'(a) < 0$ and
there exists a finite radius, $a_{min}$, such that when $a < a_{min}$,
$V(a) > 0$, and when $a \ge a_{min}$, $V(a) \le  0$. The only difference is that
in the present case, a big rip singularity develops at $a = \infty$, since $\rho \propto a^{3(|w| -1)}$, as shown in Fig. \ref{figE}.

\subsubsection{$\Lambda = 0$}

\begin{figure}
\includegraphics[width=\columnwidth]{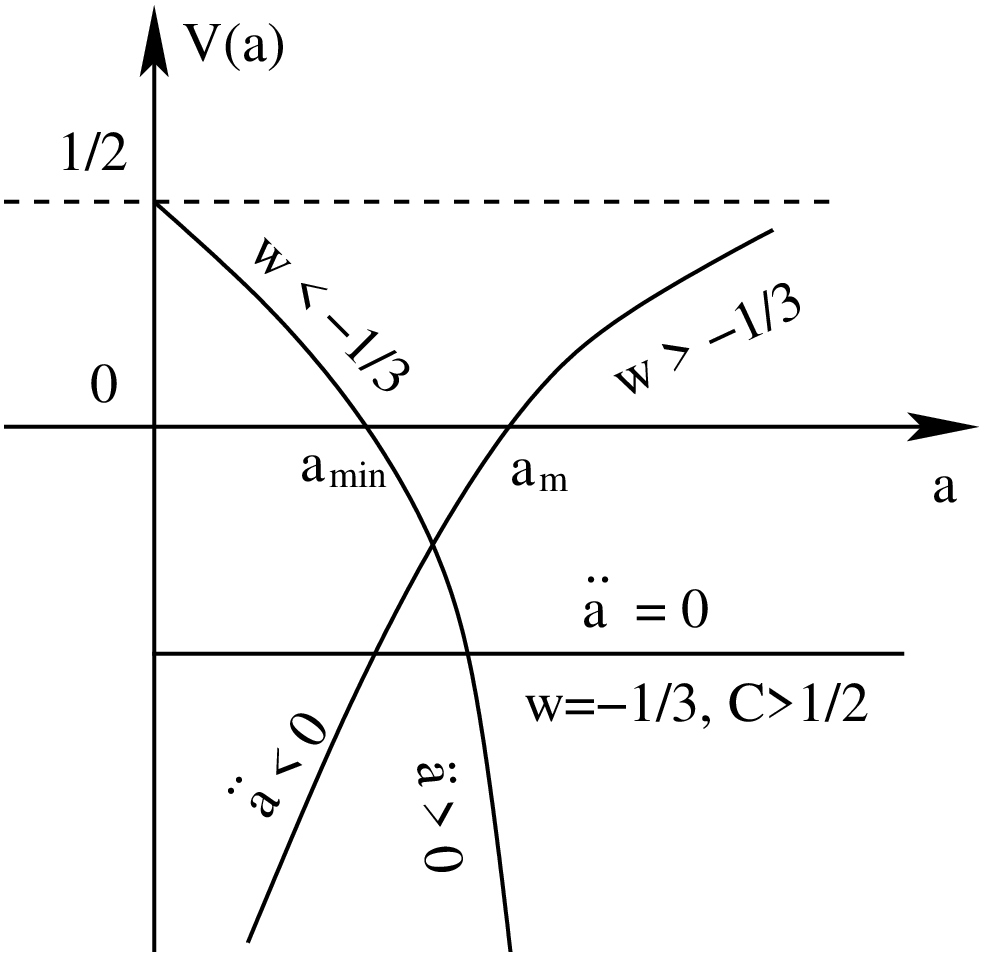}
\caption{The potential given by Eq.(\ref{2.32})  for
$k = 1 $ and $ \Lambda = 0$.}
\label{fig6}
\end{figure}

\begin{figure}
\includegraphics[width=\columnwidth]{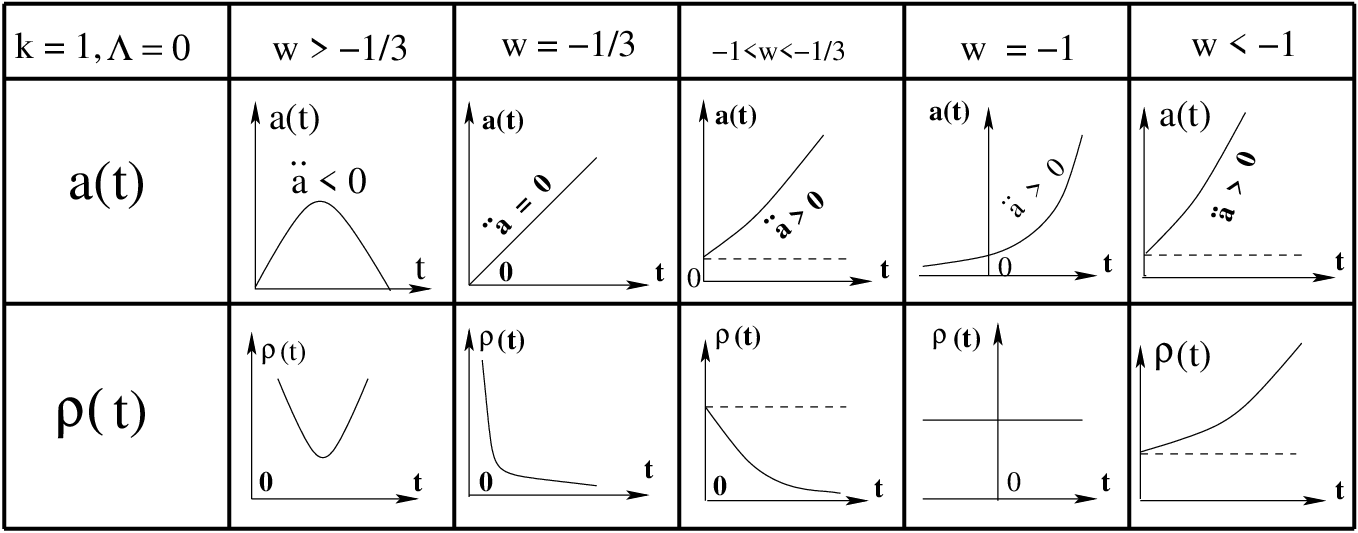}
\caption{The expansion factor $a(t)$, the acceleration $\ddot{a}(t)$, and the energy
density $\rho(t)$ for  $k = 1$ and $ \Lambda = 0$. There are both big bang
and big crunch singularities in the case  $w > -1/3$, while only a
big bang singularity occurs in the case $w = -1/3$. There is no singularity
in the cases with $- 1 \le  w <-1/3$. A big rip singularity occurs at $a = \infty$
for $w < -1$. }
\label{figF}
\end{figure}

In this case, we find that
\bq
\lb{2.32}
V(a) = \frac{1}{2}  -  \frac{{\cal{C}}}{a^{1+3w}}.
\eq
Fig. \ref{fig6} shows  the potential for various values of $w$. For $w > -1/3$, the motion of the universe is restricted
to $a \le a_{m}$, where $a_{m}$ is the solution of $V(a) = 0$. The universe starts
to expand at the big bang singularity $a = 0$ until the turning point $a = a_{m}$ is reached.
It then starts to collapse, until a big crunch singularity is developed
at $a = 0$, as shown in Fig. \ref{figF}.

When $w = -1/3$, there is motion only for ${\cal{C}} > 1/2$, for
which the universe expands linearly from a big bang singularity at
$a = 0$ with $\ddot{a} = 0$.

When $ - 1 \le w < -1/3$, motion is possible only for $a > a_{min}$, as shown
in Figs. \ref{fig6} and \ref{figF}. The universe starts to expand from the initial
point $a_{i} \ge a_{min}$ with $\ddot{a} > 0$. No turning point exists;
so, the universe will expand forever. In this case, the matter density remains
finite during the whole process, so no singularity exists.

When $w < - 1$, it can be shown that the motion for $a < a_{min}$ is also forbidden.
As a result, no big bang singularity exists; however, a big rip
singularity develops as $ a \rightarrow \infty$, as shown by Fig. \ref{figF}.
During the whole process,  $\ddot{a} > 0$.

\subsubsection{$\Lambda < 0$}

\begin{figure}
\includegraphics[width=\columnwidth]{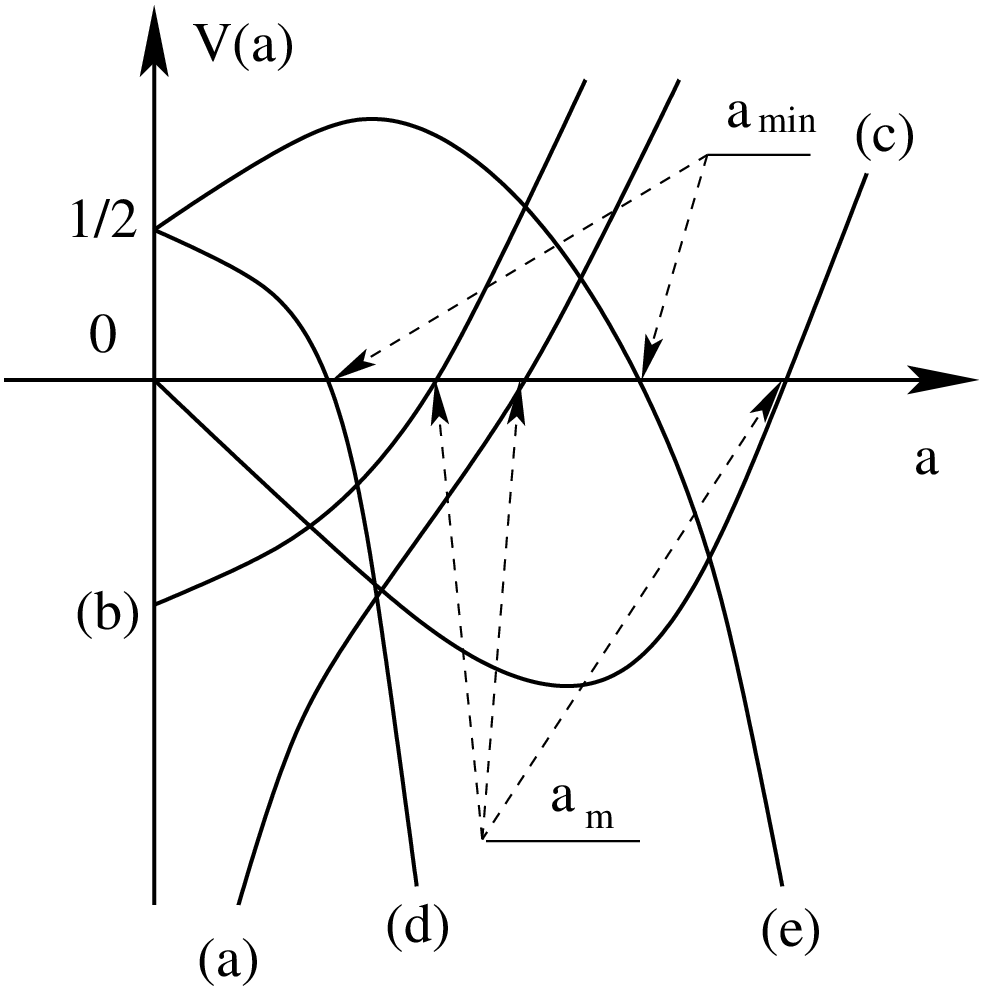}
\caption{The potential given by Eq.(\ref{2.33})  for $k = 1 $ and $
\Lambda = 0$: (a)  for $ w > -1/3$; (b)  for $w = -1/3$ and $
{\cal{C}} > 1/2$; (c) for $ - 1< w < -1/3$; (d)  for $ w  = -1$ and
$ {\cal{C}} > |\Lambda|/6 $; and (e)  for $ w < -1$.} \label{fig7}
\end{figure}

\begin{figure}
\includegraphics[width=\columnwidth]{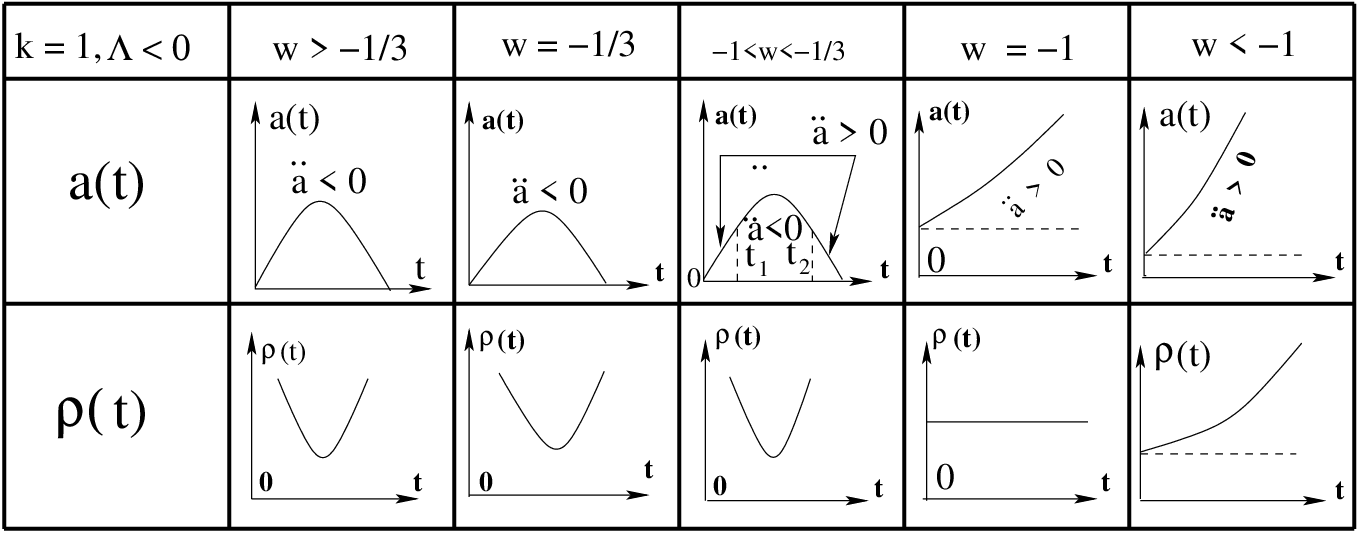}
\caption{The expansion factor $a(t)$, the acceleration $\ddot{a}(t)$, and the energy
density $\rho(t)$ for  $k = 1$ and $ \Lambda < 0$. There are both big bang
and big crunch singularities for the case $w > -1$, while only a
big bang singularity occurs for the case $w = -1$. There are  no singularities
for the case $w = -1$, while there is a   big rip singularity   at $a = \infty$
for $w < -1$. }
\label{figG}
\end{figure}

In this case, we have
\bq
\lb{2.33}
V(a) = \frac{1}{2}  + \frac{1}{6}\left|\Lambda\right| a^{2}
       -  \frac{{\cal{C}}}{a^{1+3w}},
\eq
which has the properties as shown in Fig. \ref{fig7}. In particular, when
$w > -1/3$, the universe starts to expand from a big bang
singularity at $a = 0$ until a maximal radius $a_{m}$ is reached, where $V\left(a_{m}\right)
= 0$. Afterward, the universe starts collapsing, reaching a big crunch
singularity at $t = 2 t_{m}$, where $t_{m}$ is determined by
$a_{m} = a\left(t_{m}\right)$. Throughout the process the universe is
decelerating, as shown in  Fig. \ref{figG}.

When $w = -1/3$, the potential is non-negative for ${\cal{C}} \le 1/2$, so the
motion is forbidden. When ${\cal{C}} > 1/2$ we have $V(a) < 0$ for
$a < a_{m}$, where $a_{m}$ is the root of $V(a) = 0$, as shown in Fig. \ref{fig7}.
Thus, motion is possible in the region $a < a_{m}$, for which the universe
expands from a big bang singularity at $a = 0$. Once the maximal
radius $a_{m}$ is reached, the universe collapses until a big crunch is developed at $t = 2t_{m}$.

When $- 1 < w < -1/3$, the potential is
negative only for $a < a_{m}$, as shown in Fig. \ref{fig7}. In
particular,  a big bang (crunch)
 singularity occurs at $t = 0$ ($t = 2t_{m}$).
The difference from the previous case is that there now exists
a time $t_{max}$, such that the universe is accelerating for $ 0 < t < t_{max}$ and $ 2t_{m} - t_{max} < t < 2t_{m}$, while during the time $ t_{m} - t_{max} < t
< 2t_{m} -t_{max}$ the universe is decelerating.  $t_{max}$ is the root of $V'\left(t_{max}\right) = 0$.

When $w = -1$, the potential is non-positive only  for ${\cal{C}} > |\Lambda|/6$
and $a \ge a_{min}$,  where $a_{min}$ is the root of $V(a) = 0$, as shown
in Fig. \ref{fig7}.  In this case the universe starts to expand
at an initial radius $a_{i} \ge a_{min}$, and will expand forever
with $\ddot{a} > 0$. However, spacetime is not singular even when $a = \infty$.

When $w < -1$, the potential is non-positive only  for $a \ge a_{min}$,
where $a_{min}$ is again the root of $V(a) = 0$, as shown
in Fig. \ref{fig7}. The evolution of the universe in this case is similar to the last
one, except that now a big rip singularity develops
at  $a = \infty$.

\subsection{$ k = -1$}

When $k = -1$, the potential is given by
\bq
\lb{2.34}
V(a) = - \frac{1}{2}  - \frac{1}{6}\Lambda a^{2}
       -  \frac{{\cal{C}}}{a^{1+3w}}.
\eq
To study the motion of the universe in this case, it is also convenient to
distinguish the three cases $\Lambda > 0,\; \Lambda = 0$ and $\Lambda < 0$.  In
each case there are five sub-cases for different choices of $w$.

\subsubsection{$\Lambda > 0$}

\begin{figure}
\includegraphics[width=\columnwidth]{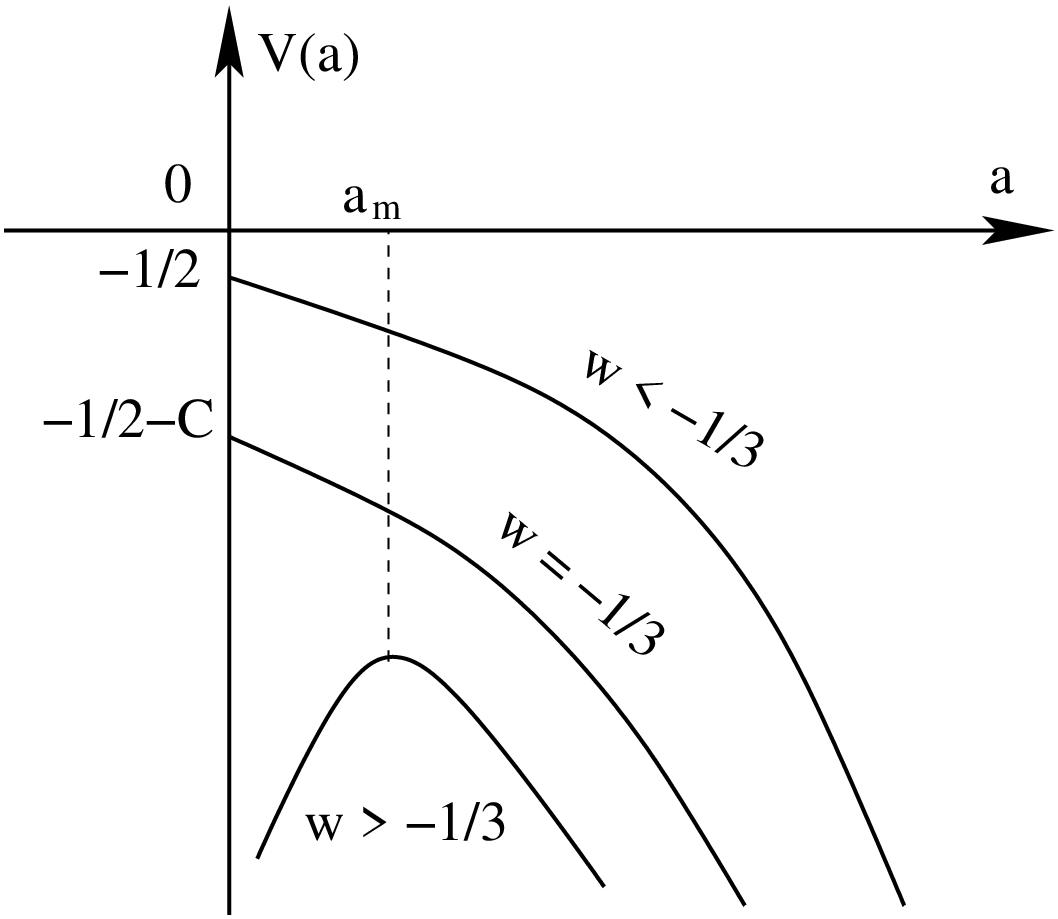}
\caption{The potential given by Eq.(\ref{2.34})  for $k = -1 $ and $ \Lambda > 0$.
}
\label{fig8}
\end{figure}

\begin{figure}
\includegraphics[width=\columnwidth]{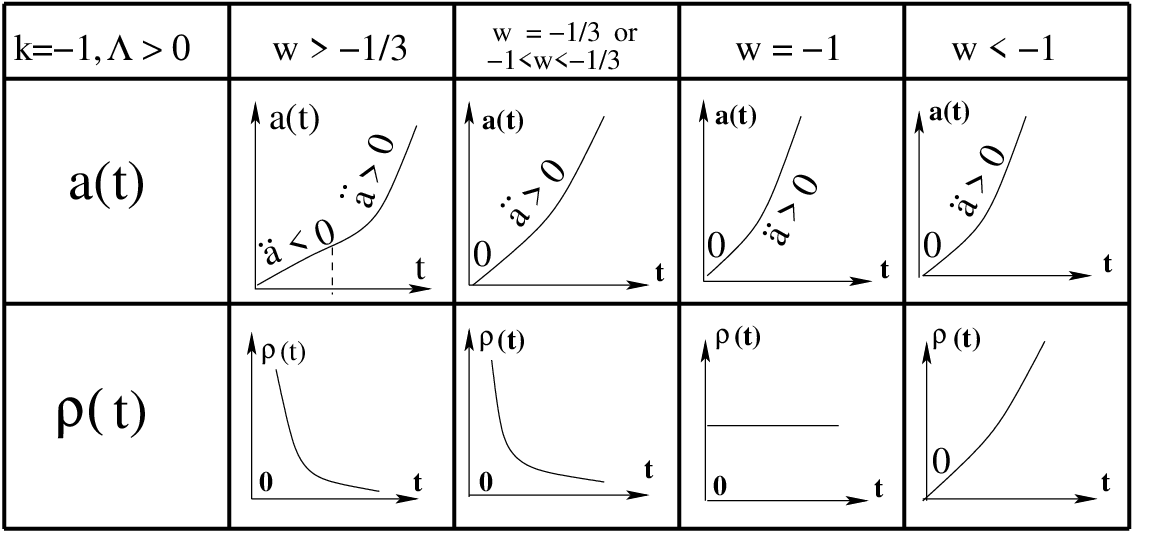}
\caption{The expansion factor $a(t)$, the acceleration $\ddot{a}(t)$, and the energy
density $\rho(t)$ for  $k = - 1$ and $ \Lambda > 0$. There is a big bang
singularity for $w > -1$, no singularity for $w = -1$, and a big rip singularity
at $a = \infty$
for $w < -1$. }
\label{figH}
\end{figure}

In this case, we find that $V(a) \rightarrow - \infty$ as $a
\rightarrow \infty$, and
 \bq \lb{2.35} 
 V(a\rightarrow \infty) = \cases{ - 1/2, & $ w <
-1/3$,\cr - (1/2 + {\cal{C}}), & $ w = -1/3$,\cr 
- \infty, & $ w >
-1/3$,\cr} 
\eq 
when $a \rightarrow 0$, as shown by Fig. \ref{fig8}.
Thus, when $w > -1/3$, the potential has a maximum at $a_{m}$ where
$V'\left(a_{m}\right) = 0$. The universe expands from a big
bang at $a = 0$. Initially, it is decelerating since $\ddot{a} < 0$.
However, after expanding to $a_{m}$, the expansion begins
accelerating, since now $\ddot{a} > 0$, as shown in Fig. \ref{figH}. When $-
1 < w \le -1/3$, the universe expands from a big bang at $a = 0$
until $a = \infty$, without a turning point since
$\ddot{a} > 0$ during the whole process. The case of $w = -1$ is similar
to the case of $- 1 < w \le -1/3$, except that spacetime is not
singular either at $a = 0$ or at $a = \infty$, as shown in
Fig.\ref{figH}. When $w < -1$, the universe starts
to expand from $a = 0$ with $\ddot{a} > 0$ for any given $a$. There
is no big bang singular at $a = 0$; however, there is a big rip
singularity at $a = \infty$.

\subsubsection{$\Lambda = 0$}

\begin{figure}
\includegraphics[width=\columnwidth]{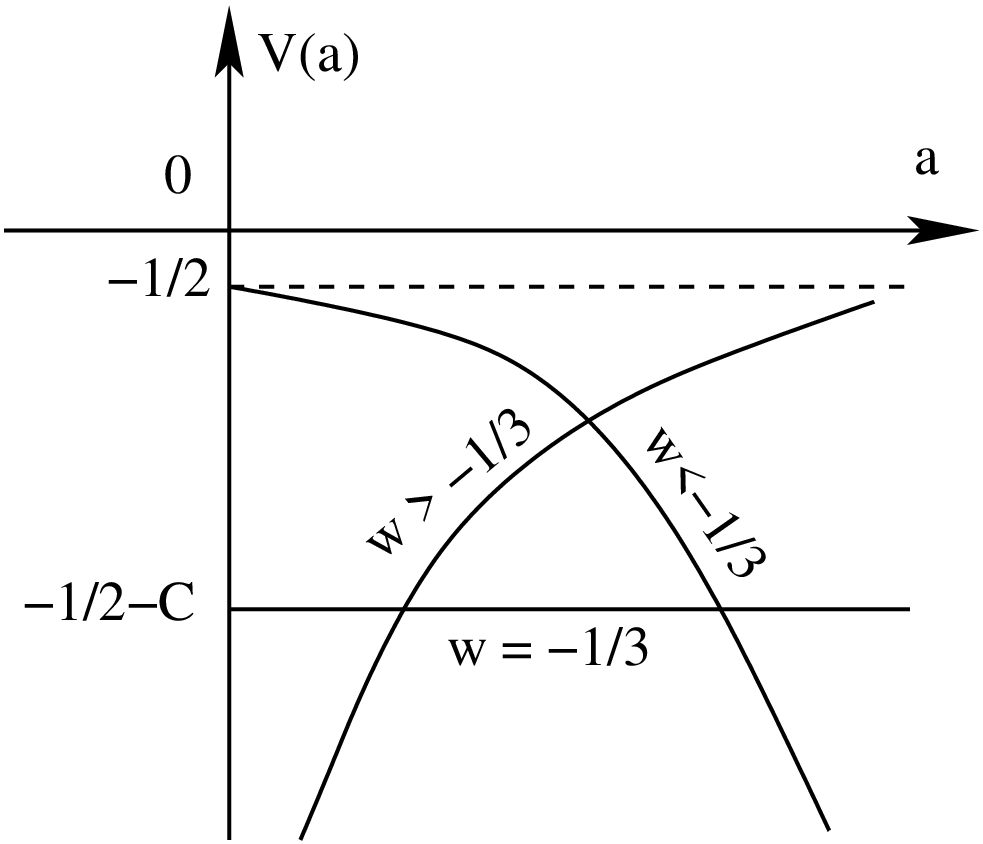}
\caption{The potential given by Eq.(\ref{2.36})  for $k = -1 $ and $ \Lambda = 0$.
}
\label{fig9}
\end{figure}

\begin{figure}
\includegraphics[width=\columnwidth]{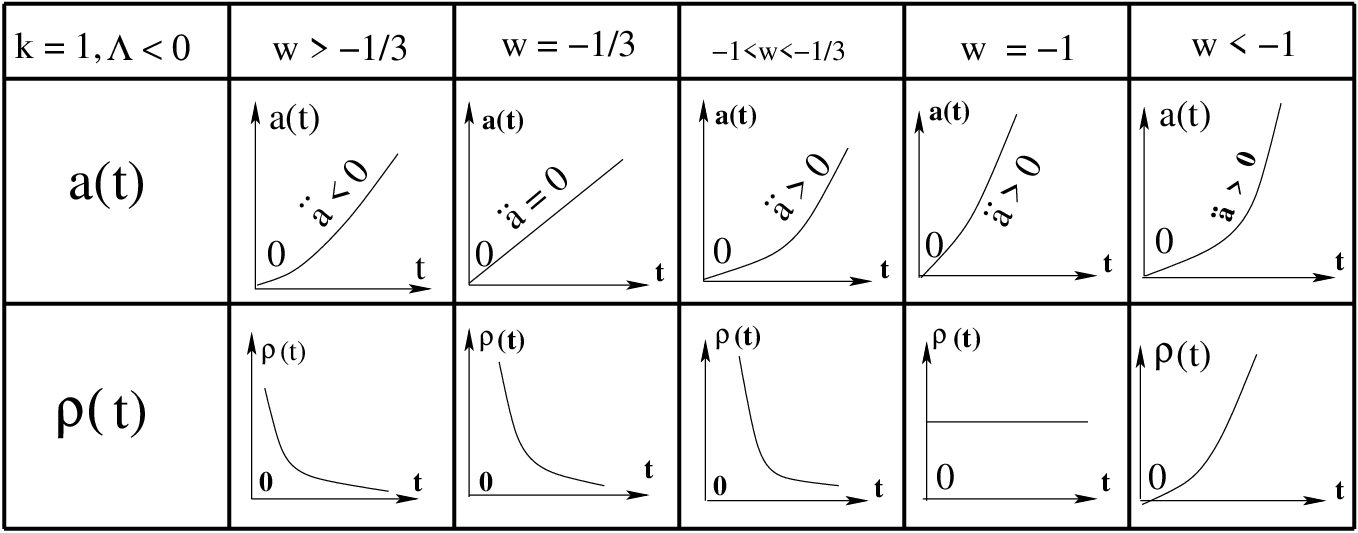}
\caption{The expansion factor $a(t)$, the acceleration $\ddot{a}(t)$, and the energy
density $\rho(t)$ for  $k = - 1$ and $ \Lambda = 0$.  There is a big bang
singularity for $w > -1$, no singularity for $w = -1$, and a big rip singularity
at $a = \infty$ for $w < -1$.}
\label{figI}
\end{figure}

In this case, we have
\bqn
\lb{2.36}
V(a) &=& - \frac{1}{2} - \frac{{\cal{C}}}{a^{1+3w}} < 0,\nb\\
\ddot{a} &=& - \frac{(3w+1){\cal{C}}}{a^{2+3w}}
= \cases{ < 0, & $ w > -1/3$,\cr
0, & $ w = -1/3$,\cr
> 0, & $ w < -1/3$,\cr}
\eqn
when $a \in [0, \infty)$, as shown by Fig. \ref{fig9}. We also have
\bq
\lb{2.37}
\rho(a) 
= \cases{ \infty, & $ w > -1/3$,\cr
\rho_{0}, & $ w = -1/3$,\cr
0, & $ w < -1/3$,\cr}
\eq
as $a \rightarrow 0$, and
\bq
\lb{2.38}
\rho(a) 
= \cases{ 0, & $ w > -1/3$,\cr
\rho_{0}, & $ w = -1/3$,\cr
\infty, & $ w < -1/3$,\cr}
\eq
as $a \rightarrow \infty$.  Fig. \ref{figI} shows the motion of the universe
for each  given $w$.

\subsubsection{$\Lambda < 0$}

In this case, we have
\bq
\lb{2.39}
V(a) =  - \frac{1}{2} + \frac{1}{6}|\Lambda|a^{2} - \frac{{\cal{C}}}{a^{1+3w}}.
\eq
Depending on the values of $\Lambda,\; {\cal{C}}$ and $w$, the potential will have
quite different properties. In the following we sstudy them case
by case.

\begin{figure}
\includegraphics[width=\columnwidth]{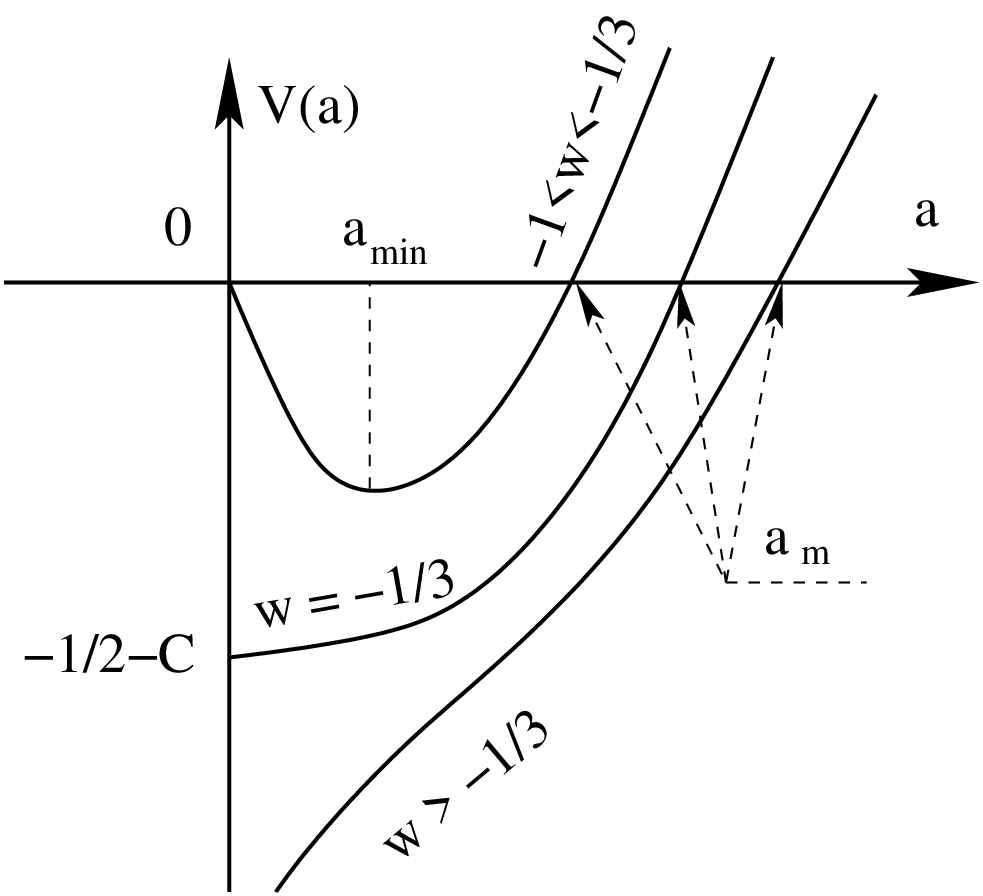}
\caption{The potential given by Eq.(\ref{2.39})  for $k = - 1,\; \Lambda < 0$
and $ w > -1$. }
\label{fig10}
\end{figure}

\begin{figure}
\includegraphics[width=\columnwidth]{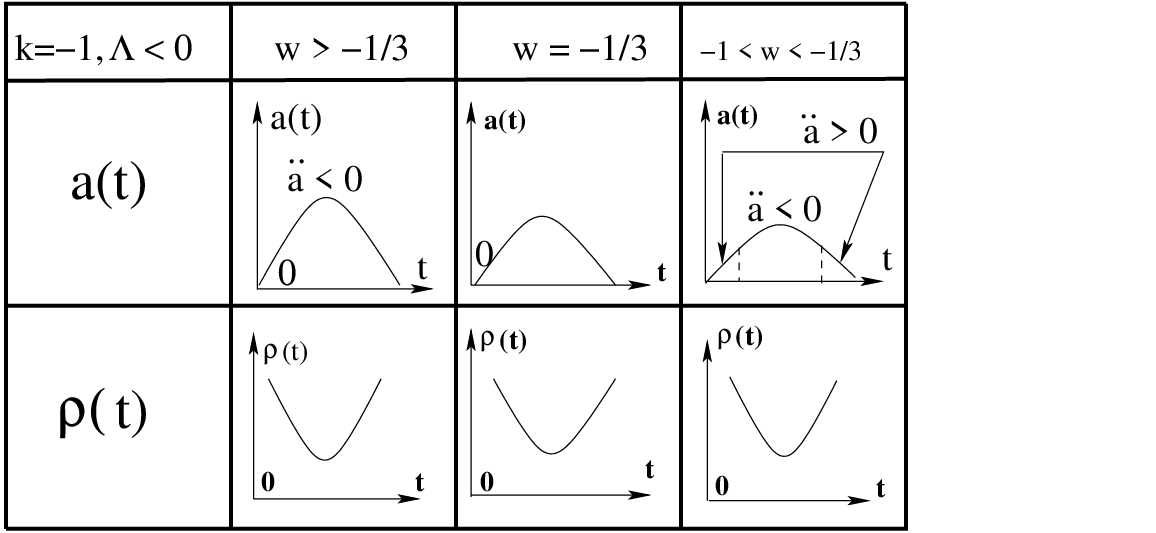}
\caption{The expansion factor $a(t)$, the acceleration $\ddot{a}(t)$, and the energy
density $\rho(t)$ for  $k = - 1,\;  \Lambda < 0$ and $w > -1$.  There are both
big bang and big crunch singularities  for all cases with $w > -1$.}
\label{figJ}
\end{figure}

{\bf Case C.3.1) $\; w > -1/3$}: In this case, the potential, shown schematically
in Fig. \ref{fig10}, is non-positive only for $a \le
a_{m}$, where $a_{m}$ is the positive root of $V(a) = 0$. Clearly, there is a big bang singularity at $a = 0$, from which the universe expands until $a = a_{m}$. Afterward, it collapses to a big crunch
singularity at $t = 2t_{m}$, for which $a\left(2t_{m}\right)
= 0$, as shown in Fig. \ref{figJ}.

{\bf Case C.3.2) $\; w = -1/3$}: The potential is similar to the
previous case, except that now $V(0) = - 1/2 - {\cal{C}}$, as shown
in   Fig. \ref{fig10}. The resulting motion of the universe is qualitatively similar to the previous case,  as shown in Fig. \ref{figJ}.

{\bf Case C.3.3) $\; - 1 < w < -1/3$}: In this case the potential has a minimum
at $a = a_{min}$, as shown in   Fig. \ref{fig10}, for which
$\ddot{a} < 0$ for $a < a_{min}$, and $\ddot{a} > 0$ for $a > a_{min}$, as
shown in Fig. \ref{figJ}.

\begin{figure}
\includegraphics[width=\columnwidth]{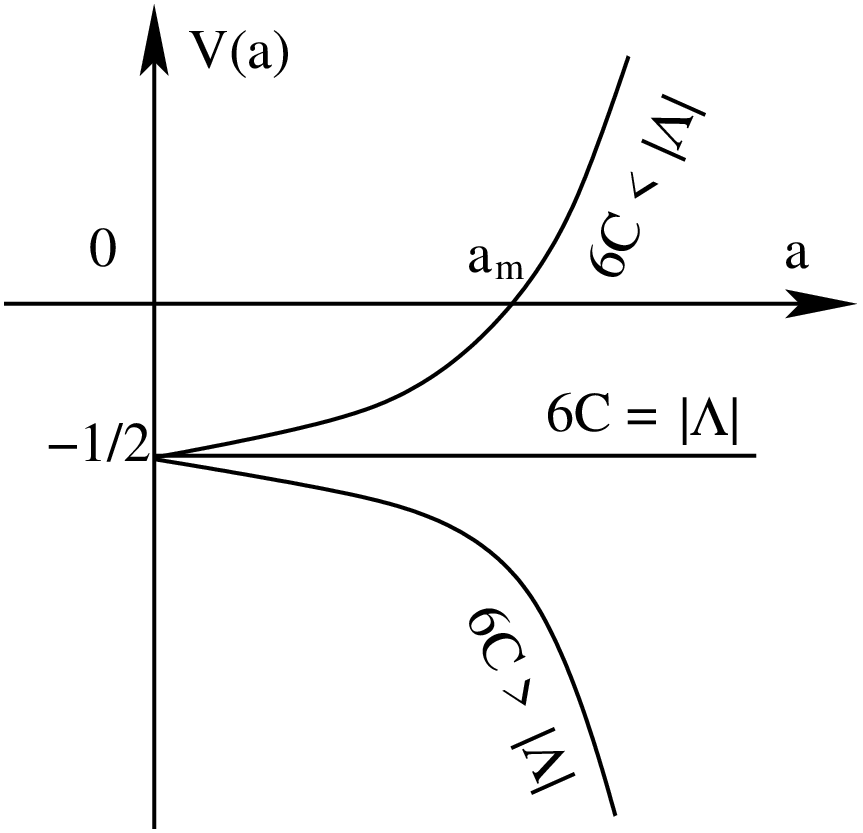}
\caption{The potential given by Eq.(\ref{2.39})  for $k = - 1,\; \Lambda < 0$
and $ w = -1$. }
\label{fig11}
\end{figure}

\begin{figure}
\includegraphics[width=\columnwidth]{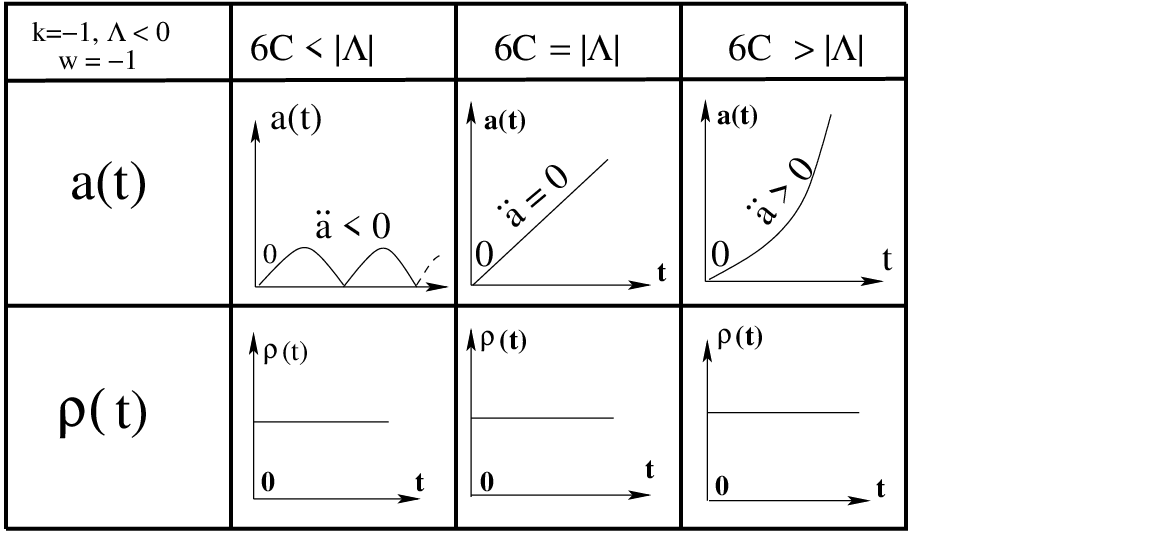}
\caption{The expansion factor $a(t)$, the acceleration $\ddot{a}(t)$, and the energy
density $\rho(t)$ for  $k = - 1,\;  \Lambda < 0$ and $w = -1$. The
spacetime is not singular in any of these cases.}
\label{figK}
\end{figure}

{\bf Case C.3.4) $\;   w  = -1$}: In this case, depending on the ratio $6{\cal{C}}
/|\Lambda|$, three sub-cases may be distinguished. When $6{\cal{C}}
/|\Lambda| < 1$, the potential is non-positive only when $a \le a_{m}$, where
$a_{m} \equiv [3/(|\Lambda| - 6{\cal{C}})]^{1/2}$, as shown in Fig. \ref{fig11}.
The universe expands from $a = 0$ until $a = a_{m}$, and then collapses until $a = 0$ again. No spacetime singularity is developed; thus, the universe oscillate between
$a = 0$ and $a = a_{m}$, as shown in Fig. \ref{figK}.

When $6{\cal{C}}/|\Lambda| = 1$, we find that $V(a) = - 1/2$, and the universe
expands linearly starting from $a = 0$. There is no turning point or
spacetime singularity, as shown in Figs. \ref{fig11} and \ref{figK}.

When $6{\cal{C}}/|\Lambda| > 1$, we find that $V(a) < - 1/2$ for any given
$a$. Starting from $a = 0$, the universe expands always at an ever accelerating rate
($\ddot{a} > 0$) until $a = \infty$, as  shown in Figs. \ref{fig11} and \ref{figK}.
No spacetime singularity is developed  during the whole process.

\begin{figure}
\includegraphics[width=\columnwidth]{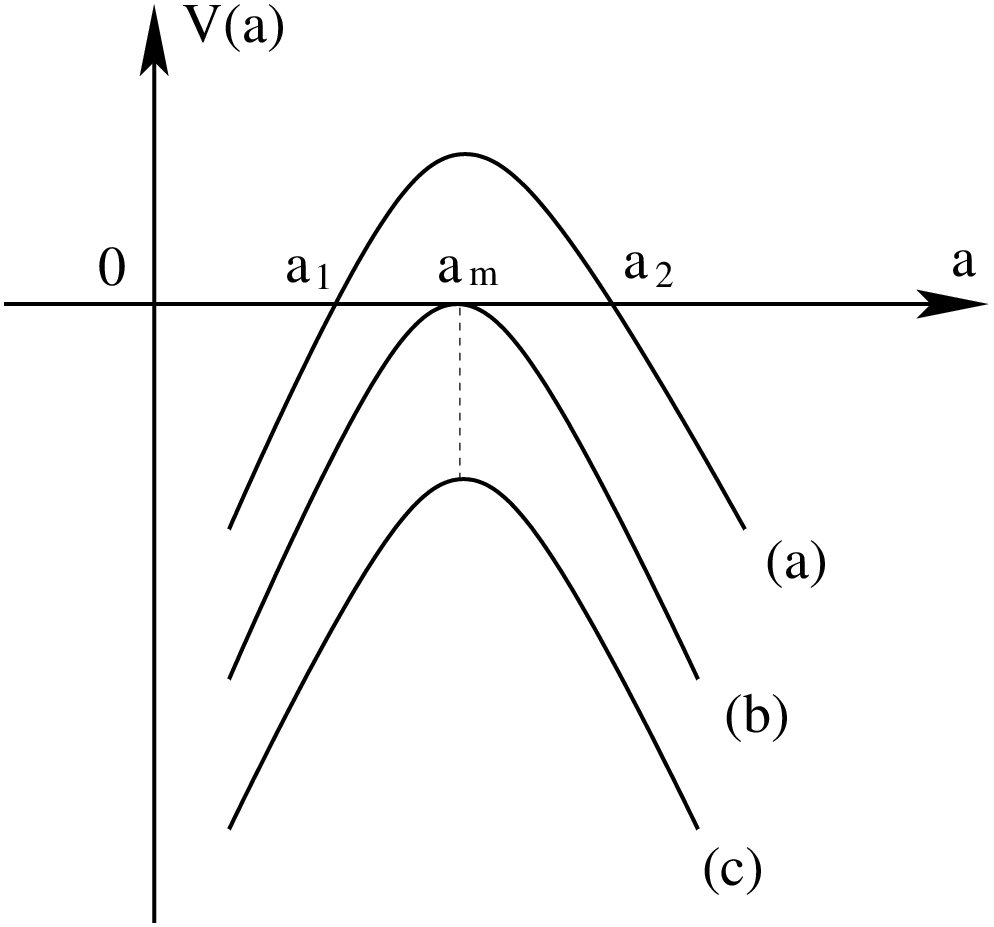}
\caption{The potential given by Eq.(\ref{2.39})
for $k = -1, \; w < - 1$ and $ \Lambda < 0$,
where $\Lambda_{c}$ is given by Eq.(\ref{2.40}). 
Curve  (a) is for $\; \left|\Lambda\right| > \left|\Lambda_{c}\right|$;
(b) is for $\; \left|\Lambda\right| = \left|\Lambda_{c}\right|$; and
(c) is for $\; \left|\Lambda\right| < \left|\Lambda_{c}\right|$.}
\label{fig12}
\end{figure}

{\bf Case C.3.5) $\;   w  < -1$}: In this case, it can be shown
that for any given  $\rho_{0}$ there exists a critical
value $\Lambda_{c}$ such that $V(a) = 0$ has two root positive roots
for $\left|\Lambda\right| > \left|\Lambda_{c}\right|$, one positive
root for $\left|\Lambda\right| = \left|\Lambda_{c}\right|$, and no
positive roots for  $\left|\Lambda\right| < \left|\Lambda_{c}\right|$,
as shown in Fig.\ref{fig12}, where $\Lambda_{c}$ is the solution of
the equations $V\left(a_{m}, \Lambda_{c}\right) = 0$, and
$V'\left(a_{m},   \Lambda_{c}\right) = 0$. It can be shown that
$\Lambda_{c}$ is given by,
\bq
\lb{2.40}
\Lambda_{c} = \left(\frac{3|w| -1}{|w| -1}\right)
\left[3\left(|w| -1\right){\cal{C}}\right]^{\frac{2}{3|w| -1}}.
\eq

\begin{figure}
\includegraphics[width=\columnwidth]{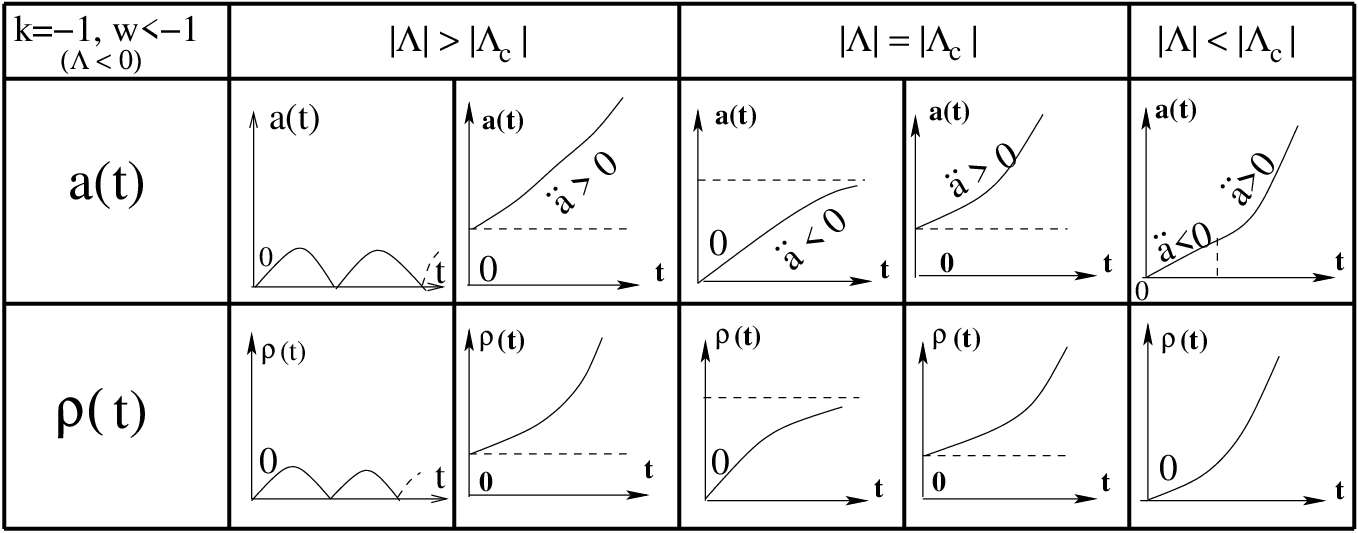}
\caption{The expansion factor $a(t)$, the acceleration $\ddot{a}(t)$, and the energy
density $\rho(t)$ for  $k = - 1, \; w < -1$ and $ \Lambda < 0$, where
$\Lambda_{c}$ is given by Eq.(\ref{2.40}). There are big rip singularities
in all cases, except when $\left|\Lambda\right| \ge
\left|\Lambda_{c}\right|$.}
\label{figL}
\end{figure}

{\bf Case C.3.5.1)    $\left|\Lambda\right| > \left|\Lambda_{c}\right|$}:
In this case,
the potential is positive in the region $a_{1} < a < a_{2}$, where
$a_{1}$ and $a_{2}$ are the two positive roots of $V(a) = 0$ with $a_{2}
> a_{1}$. Therefore,   the motion of the universe is now restricted
to the regions $ 0 \le a \le a_{1}$ and $   a \ge a_{2}$, depending on the
initial conditions. If the universe starts to expand at $a = 0$, it will
expand until reaching its maximal radius $a=a_{1}$, and then collapse
until $a= 0$. During the whole process,   $\ddot{a} < 0$.
Since  spacetime is not singular at $a = 0$, the
universe again starts to expand. This process will repeat
endlessly, as shown in Fig. \ref{figL}. However, if the universe starts to expand
at a radius $a_{i} \ge a_{2}$, it will expand forever since $\ddot{a} > 0$. A big rip
singularity finally develops, since
$\rho \rightarrow \infty$ as $a \rightarrow \infty$.

{\bf Case C.3.5.2) $\left|\Lambda\right| = \left|\Lambda_{c}\right|$}:
In this case,  there exists a static point $a_{m}$, at which
$V\left(a_{m}\right) = V'\left(a_{m}\right) = 0$, as can be seen in Fig. \ref{fig12},
where $a_{m} = \left[3\left(|w| -1\right){\cal{C}}\right]^{- {1}/(3|w| -1)}$.
Therefore, if the universe starts to expand
from   $a = 0$, it will continue to expand until $a = a_{m}$ with $\ddot{a} < 0$.
Since $\dot{a} = 0 = \ddot{a}$ at $a = a_{m}$, the universe
will become static at that point.   However,  it is not stable, and with
small perturbations, the universe will either collapse until   $a  = 0$ or expand
forever with $\ddot{a} > 0$. It should be noted that spacetime is not
singular at $a = 0$; thus, if it collapses, it begins expanding again when
the point $a =0$ is reached. If the universe initially has a radius $a_{i} > a_{m}$,
from Fig. \ref{fig12} it will continue expanding forever, since $V'(a)$ is always
negative.  A big rip singularity will ultimately develop at $a = \infty$.

{\bf Case C.3.5.3) $\left|\Lambda\right| < \left|\Lambda_{c}\right|$}:
In this case,  the potential
$V(a)$ is negative for all $a$, as shown in Fig. \ref{fig12}. Therefore,
the corresponding solutions have no turning points. If the universe initially
starts to expand from    $a = 0$, it will expand forever. However, the potential
has a maximum at $a = a_{m}$, for which
\bq
\lb{2.41}
\ddot{a} = \cases{ < 0, & $ a < a_{m}$,\cr
 = 0, & $ a = a_{m}$,\cr
  > 0, & $ a > a_{m}$.\cr}
\eq
Thus, the expansion of the universe decelerates initially; however,
the expansion begins to accelerate  once it reaches $ a_{m}$. As in
 the previous two cases,
a big rip singularity develops at $a = \infty$.

\section{Conclusions}

In this paper, we have systematically studied the solutions of the
Friedmann-Robertson-Walker (FRW) universe with a cosmological
constant and a perfect fluid having the equation of state $p =
w\rho$, where  $w$ is an  arbitrary real constant,  and $p$ and $\rho$ denote the pressure and
energy density of the fluid, respectively.
With the motion of the
universe in the form Eq.(\ref{1.1}), 
we have been able to classify all of the solutions according to the
values of $k,\; \rho_{0},\; w$ and $\Lambda$, by analogy with one-dimensional  motion in classical
mechanics \cite{GPS02}. These
solutions are classified and presented in Figs. \ref{figA},
\ref{figB}, \ref{figC}, \ref{figD}, \ref{figE}, \ref{figF},
\ref{figG}, \ref{figH}, \ref{figI}, \ref{figJ}, \ref{figK}, and
\ref{figL}. Such classifications are unique, as the potential $V(a)$ depends on these parameters, and different choices of them
will correspond to different regions in the phase space,  and consequently to different motions of the universe.  
Some particular cases have previously been discussed in  various
standard textbooks.  In some cases, we have corrected errors in classification and interpretation.
 
In this paper, we have studied  the FRW universe in the Einstein theory of
gravity. As pointed out in Introduction, such studies can easily be applied to all models in which
the evolution of the universe can be cast in the form of Eq.(\ref{1.1})
\cite{DEs}, including brane worlds in string/M theory \cite{Sbranes},
asymmetric branes \cite{Pad05}, and cosmological models in the Horava-Lifshitz 
gravity \cite{HL09}.

\begin{acknowledgments}
We would like very much to express our gratitude to Brandon Bear for his critical reading of this
paper and valuable suggestions.  AW  is supported in part by DOE Grant, DE-FG02-10ER41692
 
\end{acknowledgments}


\begin{thebibliography}{nbound}


\bibitem{agr98} A.G. Riess  {\it et al.}, Astron. J. {\bf 116}, 1009 (1998);
  S. Perlmutter {\it et al.}, Astrophys. J. {\bf 517}, 565 (1999).

\bibitem{Obs} A.G. Riess {\it et al.}, Astrophys. J. {\bf 607}, 665 (2004);
   P. Astier {\it et al.}, Astron. and Astrophys. {\bf 447}, 31 (2006);
  D.N. Spergel {\it et al.}, astro-ph/0603449;
  W.M. Wood-Vasey {\it et al.}, astro-ph/0701041;
  T.M. Davis {\it et al.}, astro-ph/0701510.


\bibitem{DEs} V. Sahni  and A. A. Starobinsky, Int. J. Mod. Phys. D
{\bf 9}, 373 (2000);  P.J.E. Peebles and B. Ratra, Rev. Mod. Phys {\bf 75}, 559 (2003);
T. Padmanabhan, Phys. Rep. {\bf 380}, 235 (2003);
V. Sahni, ``{\em Dark Matter and Dark Energy}," arXiv:astro-ph/0403324 (2004);
{\it The Physics of the Early Universe}, edited by
E. Papantonopoulos (Springer, New York 2005), P. 141;
T. Padmanabhan, Proc. of the 29th Int. Cosmic Ray Conf. {\bf 10}, 47 (2005);
E.J. Copeland, M. Sami, and S. Tsujikawa, ``{\em Dynamics of dark energy},"
arXiv:hep-th/0603057 (2006); E.W. Kolb, ``{\em Cosmology and the Unexpected},"
arXiv:0709.3102; E. Linder, ``{\em Mapping the cosmological expansion},"
arXiv:0801.296; J.A. Frieman, M.S. Turner, and D. Huterer, ``{\em
Dark Energy and the Accelerating Universe}," arXiv:0803.0982.



\bibitem{DETF} A. Albrecht, {\em et al.}, ``{\em Report of the dark energy
task force}," arXiv:astro-ph/0609591.


\bibitem{SCH07} S. Sullivan, A. Cooray, and D.E. Holz, arXiv:0706.3730;
A. Mantz, {\em et al.},  arXiv:0709.4294.


\bibitem{Ch07} A.D. Chernin, {\it et al.}, arXiv:0704.2753;  arXiv:0706.4171.


\bibitem{wen} S. Weinberg, Rev. Mod. Phys. {\bf 61}, 1 (1989);
  S.M. Carroll, arXiv:astro-ph/0004075;
   T. Padmanabhan, Phys. Rept. {\bf 380}, 235 (2003);
   S. Nobbenhuis,  arXiv:gr-qc/0411093;
   J. Polchinski,  arXiv:hep-th/0603249.
   J..M. Cline, arXiv:hep-th/0612129.

\bibitem{Fish} W. Fischler, {\em et al.}, JHEP, {\bf 07}, 003 (2001);
J.M. Cline, {\em ibid.}, {\bf 08}, 035 (2001);
E. Halyo, {\em ibid.}, {\bf 10}, 025 (2001); S. Hellerman,
{\em ibid.}, {\bf 06}, 003 (2003).

\bibitem{KS00}
 L.M. Krauss and R.J. Scherrer,   arXiv:0704.0221; and references therein.


\bibitem{quit}
 R.~R.~Caldwell, R.~Dave and P.~J.~Steinhardt, Phys. Rev. Lett. {\bf 80}, 1582 (1998);
 A.~R.~Liddle and R.~J.~Scherrer, Phys. Rev. D{\bf 59}, 023509 (1999);
 P.~J.~Steinhardt, L.~M.~Wang and I.~Zlatev, {\em ibid.}, {\bf 59}, 123504 (1999).


\bibitem{DGP00}  G.~R.~Dvali, G.~Gabadadze and M.~Porrati,
  Phys. Lett. B{\bf 484}, 112 (2000);
  C.~Deffayet, {\em ibid.}, {\bf 502}, 199 (2001); V. Sahni and Y. Shtanov,
     JCAP, {\bf  0311}, 014 (2003).

\bibitem{FR}
  S. Capozziello, S. Carloni, and A. Troisi, arXiv:astro-ph/0303041;
  S.M. Carroll, et al, Phys. Rev. D{\bf 70}, 043528 (2003);
  S. Nojiri and S.D. Odintsov, {\em ibid.}, {\bf 68}, 123512 (2003).

\bibitem{GPS02} H. Goldstein, C. Poole, and J. Safko, Classical Mechanics, Third edition
(Addison Wesley, New York, 2002).

\bibitem{szy}  M. Szydlowski and W. Czaja, PRD {\bf 69},  083507; 083518 (2004); Ann. Phys. 
 {\bf 320}, 261  (2005); M.Szydlowski, Int. J. Mod. Phys. A {\bf 20}, 2443 (2005); 
  M. Szydlowski and O.Hrycyna, Gen. Relativ. Grav. {\bf 38} 121 (2006); 
  M. Szydlowski and O.Hrycyna, and A. Krawiec, JCAP, {\bf 06}, 010 (2007);
  and M. Szydlowski and O.Hrycyna, J. Math. Phys. {\bf 49}, 072703 (2008).


\bibitem{CTS05} T. Chiba, R. Takahashi, and N. Sugiyama, Class. Quantum Grav. {\bf 22},
3745 (2005).


\bibitem{d'Inverno}  R. d'Inverno, Introducing Einstein's Relativity (Clarendon Press,
Oxford, 2003).

 \bibitem{HE72} S. Hawking and G.F.R. Ellis, The Large Scale Structure of Space-Time (Cambridge University Press, 1972).

 \bibitem{SD03} S. Dodelson,  Modern Cosmology (Academic  Press, New York, 2003).


\bibitem{Sbranes} Y.-G.   Gong, A. Wang, and Q. Wu,   Phys. Lett. B{\bf 663}, 147
     (2008) [arXiv:0711.1597]; and A. Wang and N.O. Santos, Phys. Lett. B {\bf 669}, 127 (2008)
     [arXiv:0712.3938]; arXiv:0808.2055;   Q. Wu, N.O. Santos, P. Vo,  and A. Wang,  JCAP, 
     {\bf 09} 004 (2008) [arXiv:0804.0620]; Q. Wu, Y. Gong,  and A. Wang,    JCAP, 06, 015 (2009)
     [arXiv:0810.5377 ].


\bibitem{Pad05} A. Padilla, Class. Quantum Grav. {\bf 22}, 681 (2005); {\bf 22}, 1087
(2005); K. Koyama and K. Koyama, Phys. Rev. D{\bf 72}, 043511 (2005);
C. Charmousis, R. Gregory, and A. Padilla, arXiv:0706.0857; Y. Shtanov, {\em et
al}, arXiv:0901.3074; and references therein.

\bibitem{HL09} P. Horava, JHEP, {\bf 03}, 020 (2009); Phys. Rev. D{\bf 79}, 084008 (2009);
Phys. Rev. Lett. {\bf 102}, 161301 (2009); 
T. takahashi and J. Soda, arXiv:0904.0554;
G. Calcagni, arXiv:0904.0829;
E. Kiritsis and G. Kofinas, arXiv:0904.1334;
H. L\"u, J. Mei, and C.N. Pope, arXiv:0904.1595; 
S. Mukohyama, arXiv:0904.2190;
R. Brandenberger, arXiv:0904.2835;
A. Volovich and C. Wen, arXiv:0904.2455;
R.-G. Cai, Y. Liu, and Y.-W. Sun, arXiv:0904.4104;
X. Gao, arXiv:0904.4187; 
E. Colgain and H. Yavartanoo, arXiv:0904.4357;
A. Wang and Y. Wu, JCAP, {\bf 07}, 012 (2009) [arXiv:0905.4117];
A. Wang and R. Maartens, arXiv:0907.1748;
 and references therein.

\end{thebibliography}
\end{document}